\documentclass[11pt]{article}
\usepackage[margin=1in]{geometry}
\usepackage[T1]{fontenc}
\usepackage[utf8]{inputenc}
\usepackage{lmodern}
\usepackage{microtype}
\usepackage{setspace}
\usepackage{booktabs}
\usepackage{graphicx}
\usepackage{array}
\usepackage{amsmath}
\usepackage{enumitem}
\usepackage{float}
\usepackage{placeins}
\usepackage{pdflscape}
\usepackage[authoryear,round]{natbib}

\usepackage[hidelinks]{hyperref}
\usepackage{xurl}
\usepackage{caption}
\graphicspath{{figures/}}
\setlist{nosep}

\DeclareUnicodeCharacter{2212}{\ensuremath{-}}
\DeclareUnicodeCharacter{2248}{\ensuremath{\approx}}
\DeclareUnicodeCharacter{2264}{\ensuremath{\leq}}
\DeclareUnicodeCharacter{2265}{\ensuremath{\geq}}
\DeclareUnicodeCharacter{00D7}{\ensuremath{\times}}
\DeclareUnicodeCharacter{00B1}{\ensuremath{\pm}}
\DeclareUnicodeCharacter{03C1}{\ensuremath{\rho}}

\title{\textbf{Generative AI Produces and Transmits Beauty Bias}}

\author{
AJ Alvero\textsuperscript{1}\thanks{Correspondence: \href{mailto:ajalvero@cornell.edu}{ajalvero@cornell.edu}. \\ We thank Aditya Gulati, Louis Boucherie, and Wilma Bainbridge for sharing data. We thank Ben Domingue, Marissa Thompson, Sam Trejo, Austin van Loon, anthony antonio, Farnam Mohebi, and Leon Marbach for helpful comments and feedback, along with participants in SICSS-UDC, the ASA Midyear Methodological Conference, the Idealized Bodies session at the ASA Annual Meeting, the International Roundtable on Computational Social Science at the Institute for Analytical Sociology, Linköping University, the ICQCM Summit, the ICLR Workshop on Human-AI Coevolution, and the Center for Data Science for Enterprise and Society workshop.} \and
Deepak Varuvel Dennison\textsuperscript{1} \and
Alexandra Crespin\textsuperscript{2} \and
Heeyon Kim\textsuperscript{1} \and
Alejandra Regla-Vargas\textsuperscript{3} \and
Avani Akerkar\textsuperscript{1} \and
Jasmine McNealy\textsuperscript{2} \and
Ren\'e Kizilcec\textsuperscript{1}
}

\date{}

\begin{document}
\maketitle

\begin{center}
\small
\textsuperscript{1}Cornell University, Ithaca, NY, USA\\
\textsuperscript{2}University of Florida, Gainesville, FL, USA\\
\textsuperscript{3}University of Pennsylvania, Philadelphia, PA, USA\\
\end{center}

\newcommand{\SCUTWBMthree}{+0.472}
\newcommand{\SCUTWBRemain}{77}
\newcommand{\SCUTWNMthree}{+0.593}
\newcommand{\SCUTWNRemain}{77}
\newcommand{\MEWBMthree}{+0.737}
\newcommand{\MEWBRemain}{99}
\newcommand{\MEWNMthree}{+0.885}
\newcommand{\MEWNRemain}{88}
\newcommand{\SwinWBMthree}{0.000}
\newcommand{\SwinWBRemain}{0}
\newcommand{\SwinWNMthree}{-0.005}
\newcommand{\SwinWNRemain}{-3}

\begin{abstract}
Societies reward people judged to be attractive by standards intertwined with gender and race. In this paper, we examine if and how generative AI may contribute to these dynamics. We audited 27 text-to-image generators released between 2022 and 2026 using simple prompts that never mention attractiveness or beauty, analyzing 51,300 generated faces and 160,196 photographs of real people across primary and supplementary comparisons. We used three automated measures and collected ratings from 550 human raters on a subset of images. Nearly all generators produced faces scoring above photographed faces under automated measures, and human raters judged the outputs of all 23 generators as significantly more attractive than photographed faces, even after correction for multiple comparisons. When focusing our analysis on gender, we find that 37\% of female-prompted generated faces outscored the median professional fashion model. Across racial, ethnic, and national categories, groups associated with darker skin tones tended to receive lower attractiveness scores than groups associated with lighter skin tones. In a preregistered hiring simulation experiment (N = 570), exposure to AI-generated faces selected for high human-rated attractiveness lowered hireability and trustworthiness ratings and reduced interview advancement for an identical test set of candidates paired with photographs of real people; exposure to retouched photographs produced no such effect. Generative AI can reproduce racial hierarchies and gendered patterns of beauty and alter the context in which real people are judged.
\end{abstract}

\pagebreak

\section*{Introduction}
People perceived as having attractive faces earn more, are judged more competent and trustworthy, and receive more favorable treatment in settings ranging from hiring to criminal sentencing \citep{hamermesh1994beauty,dion1972beautiful,langlois2000maxims,wilson2015face,stewart1980defendant}. Beyond aptitude or qualifications, attractiveness is a valuable resource in evaluative contexts that reflects social norms and standards that are defined by people and institutions. Employers recruit and cultivate workers who ``look the part'' \citep{warhurst2000aesthetic}, fashion agents and clients decide which models' looks are marketable and what they are worth \citep{mears2011pricing}, and these judgments systematically value some racial groups, skin tones, and genders over others \citep{collins2000black,hunter2005race,hunter2007persistent,monk2015cost,dixon2017skin,monk2021beholding,taddess2026withinrace}. We therefore treat beauty not as an intrinsic property of a face but as an evaluative judgment produced through socially situated standards of attractiveness that circulate through many of the same channels from which generative AI models draw their training data. Those standards can vary across evaluators, institutions, and measurement systems while still producing patterned agreement about whose appearances are more or less valued. Much research on beauty focuses on the consequences of being judged more or less attractive, but less attention has been paid to how the faces available for evaluation come to be selected, edited, or made.

Generative artificial intelligence intervenes in this process. Earlier visual industries selected among existing people, edited their appearances, or curated which images audiences saw. Text-to-image systems instead synthesize the represented person: a prompt for a doctor, a retail worker, a Nigerian woman, or simply ``a person'' yields a face whose age, skin tone, morphology, styling, and expression are jointly constructed by the model and the humans building them at a marginal cost that makes it feasible to populate advertisements, websites, training materials, and stock imagery with people who do not exist. These tools are also widely used, and exposure to generated images is becoming increasingly common as they enter the broader visual media environment through everyday use of the internet and basic curiosity about image generation models. Their accessibility also makes image generation available to users without specialized technical expertise.

Prior research on AI bias has documented demographic disparities across modalities. In language models, racialized linguistic cues can shape judgments even when race is not explicitly stated: models respond more negatively to African American English and make less favorable judgments about speakers’ employment and criminality \citep{hofmann2024dialect}. In visual systems, audits have documented demographic disparities in classification and recognition \citep{buolamwini2018gender,raji2020saving}, while online images can amplify gender bias relative to text and shape viewers’ beliefs \citep{guilbeault2024online}. Text-to-image generation raises a related question: how inequality appears in the representations of people that systems produce. Audits of generators have found occupational gender and racial stereotypes, associations between darker skin and lower-status contexts, and homogenization within demographic groups \citep{bianchi2023easily,naik2023social,luccioni2023stable,sun2024bias,aldahoul2025faces}, while the aesthetic scoring models used to filter training data rate images of some demographic groups as higher quality than others \citep{taylor2026gaze}. Attractiveness itself, however, has received less direct attention as an outcome. Prior work has examined how models link attractive faces to favorable traits \citep{gulati2024beautiful,doh2025algorithms,doh2026aesthetics}, but less is known about the distributions of attractiveness produced by image generators when beauty is never specified in the prompt. This omission matters because users generating images of people are unlikely to specify attractiveness in many ordinary prompts, instead asking simply for a person or a social category such as a teacher, doctor, or woman. The simplicity of these prompts allows us to examine the appearances models supply when asked to render people and social categories, rather than appearances elicited by explicit aesthetic instructions.

These distributions raise three questions. The first concerns the overall attractiveness of generated people relative to real ones. Contemporary generators are trained on large visual corpora and then shaped by preference tuning, filtering, and other product-level optimization favoring outputs users find visually appealing, which gives reason to expect generated faces to be more attractive than photographed comparison populations. The second concerns how aesthetic value is distributed across social categories. For example, work on colorism and racialized beauty shows that evaluators can disagree about who is most attractive while reproducing common inequalities in whose appearance is valued least \citep{collins2000black,hunter2005race,hunter2007persistent,monk2015cost,dixon2017skin,monk2021beholding,taddess2026withinrace}. Standards of attractiveness are also gendered, with women’s and men’s appearances subject to different expectations and forms of evaluation \citep{mears2011pricing,guilbeault2024online}. These patterns underscore that beauty is not a fixed property of a face or a single underlying scale waiting to be measured. We therefore treat computational and human attractiveness ratings as measures of particular evaluative standards rather than as ground-truth labels of beauty. Generative AI raises the stakes because these systems can produce and circulate human representations at enormous scale. Many draw on overlapping web-scale visual corpora, are optimized toward outputs users find appealing, and repeatedly render the same racial and gender categories across countless applications and contexts. As generated people increasingly populate advertisements, websites, workplaces, educational materials, and other parts of the visual environment, the question is whether these systems reproduce, amplify, or reorganize existing standards of attractiveness.

If generated faces are both more attractive than real ones and unequally so, a third question follows: what are the consequences for how real people are judged? Exposure to highly selected or edited imagery alters appearance comparison and subsequent judgment \citep{fardouly2015social,holland2016systematic,engeln2020compared}, and classic contrast experiments show that exposure to highly attractive people can lower subsequent attractiveness judgments \citep{kenrick1980contrast,kenrick1989erotica}. Generative AI can deliver repeated streams of distinct faces shaped by the same aesthetic processes. Whether exposure to AI-generated faces alters consequential evaluations of real people remains unclear.

To address the first two questions, we audit 27 text-to-image systems released between 2022 and 2026 (Study 1a). Every system received the same prompts, crossing 14 racial, ethnic, and national categories, two sex terms, and eight occupations. We compare the resulting 51,300 faces with 89,435 photographs of real people using three computational attractiveness models trained on ratings from distinct human populations. To test whether human perceptions match our automated attractiveness measures, we had 550 U.S. adults rate a stratified subset of 1,374 generated and photographed faces (Study 1b). Finally, we address our third question with a hiring experiment in which 570 U.S. adults evaluated candidates paired with unmodified real, retouched, or AI-generated faces, separating the direct effect of those pairings from their downstream effect on evaluations of identical real candidates (Study 2). All primary analyses were preregistered (\url{https://osf.io/2az9q}).

\section*{Results}
\subsection*{Generated faces are more attractive than photographs of real people}

The 27 audited systems span the major proprietary consumer platforms (OpenAI, Google, Midjourney, xAI), the most widely used open-weight model families (Stable Diffusion, FLUX), and open-weight models from six other developers (Supplementary Materials). Our two primary automated measures are neural networks trained to predict human attractiveness ratings in two benchmark datasets, SCUT-FBP5500~\citep{liang2018scut} and MEBeauty~\citep{lebedeva2021mebeauty}. Under both primary automated measures, 24 of the 27 systems produce faces that score significantly above photographs of real people. A third measure, CelebA/SwinFace~\citep{qin2024swinface,liu2015celeba}, comes from a model that labels facial attributes such as "attractive" in celebrity photos; because it was not built specifically to rate attractiveness, we treat it as a robustness check. This measure also places generated faces above photographs when pooled across systems (Fig.~1A; Table~S1). Averaging across image generators, generated faces score 1.23 standard deviations (SD) above photographs under SCUT, 0.94 SD under MEBeauty, and 0.43 SD under CelebA/SwinFace. Our prompts did not mention attractiveness, beauty, or appearance. All three automated measures also clearly separate faces generated from prompts that explicitly requested attractive versus unattractive people \citep{doh2026aesthetics}, confirming that they detect attractiveness differences in synthetic images (Supplementary Materials).

We use “beauty bias” to refer to systematic differences in who generative systems make attractive, and to the consequences those differences can have for how real people are evaluated. The magnitude of the generated--photographed attractiveness gap varies across systems (Fig.~1A). Averaged across the three automated measures, 13 of the 27 systems exceed photographs by more than one SD; most are earlier or open-weight models, while three recent commercial systems (Nano Banana Pro, Grok Image, and Nano Banana 2) show no gap or a negative one. On SCUT, the nine consumer-platform systems average 0.72 SD, compared with 1.49 SD for the 18 open-weight systems. Where a model family has clearly ordered releases, the gap shrinks over time: from $d=0.88$ (v6) to 0.77 (v7) and 0.36 (v8) for Midjourney, and from 1.15 (GPT Image 1.5) to 0.65 (GPT Image 2) for OpenAI (Table~S1). Model releases differ in training data, preference tuning, filtering, and architecture all at once, which limits our ability to identify which change drove the decline in the generated--photographed attractiveness gap, or whether human raters perceive the same decline.


Human ratings do not show the same decline (Fig.~1C). In Study 1b, 550 U.S. adults rated a preregistered stratified subset of 1,374 generated and photographed faces on a 1--7 scale. Each of the 23 generators with at least 10 rated images was rated more attractive than photographs (mean 3.10), by 0.34 to 2.06 points on the scale; all differences remained significant after Holm correction (Table~S11). Two other generators, Stable Diffusion 1.5 and 2.1, were also represented in the human-rating study but contributed only four rated images each and were excluded from the comparisons because of the limited precision of those estimates. This pattern held within each of the six broad racial and ethnic categories (Fig.~2B) and also for the three systems with no gap in automated measures. Across releases, human-rated gaps changed little for OpenAI (+2.06 to +1.81 points) and rose for Midjourney (+0.97, +1.57, and +1.62 for v6 to v8), the opposite of the automated trend. The automated measures thus indicate smaller generated–photographed attractiveness gaps than human evaluators do for several recent systems. Because the automated measures correlate only moderately with human ratings ($r=0.48$ to $0.61$), we use them to compare patterns across systems and demographic groups, but rely on human ratings for claims about how attractive people find the faces.

\begin{figure}[p]
\centering
\includegraphics[width=\textwidth]{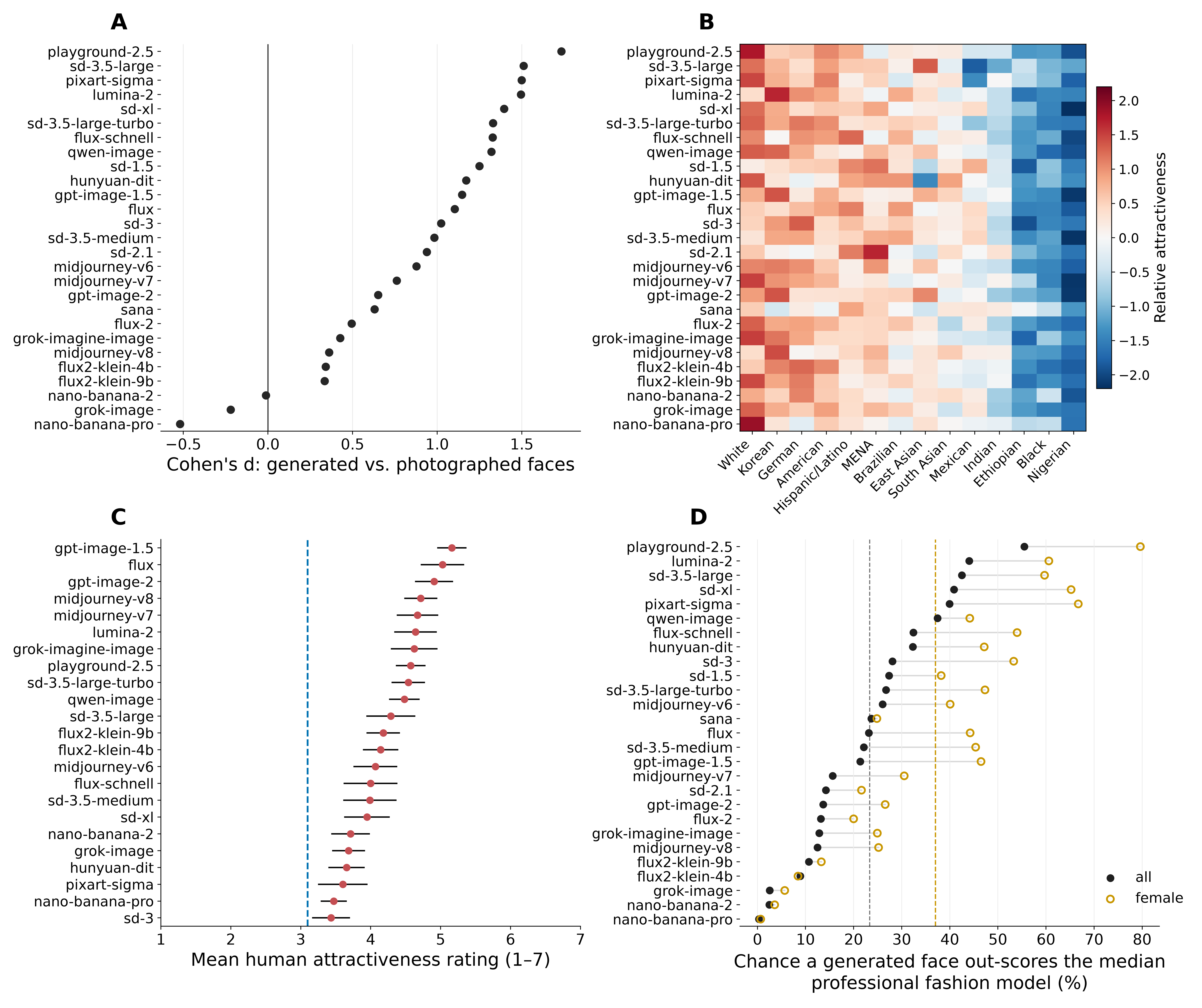}
\caption{\textbf{Generated faces are generally more attractive than photographs across automated and human evaluations.}
\textbf{A}, Generated--photographed attractiveness difference for each of 27 image generators, expressed as the mean Cohen's $d$ across the three automated attractiveness instruments.
\textbf{B}, Standardized attractiveness of each prompted demographic category within each generator. Values are standardized within generator and instrument and averaged across the three automated instruments; rows follow the generator ordering in panel A.
\textbf{C}, Mean human attractiveness rating (1--7) of AI-generated faces by generator in the human-rated subset of the audit. The panel shows the 23 generators represented by at least 10 rated images; the photographed comparison mean is shown as a reference and error bars indicate 95\% confidence intervals around generator image means.
\textbf{D}, Probability that a generated face scores above the median professional fashion model, averaged across the two primary automated attractiveness measures (SCUT and MEBeauty). Filled points show all generated faces and open points female-prompted faces. The gray dashed line marks the pooled probability across all generated faces (23\%), and the gold dashed line marks the pooled probability among female-prompted generated faces (37\%); generators are ordered by the all-face probability in this panel.}
\label{fig:main1}
\end{figure}

To gauge the extent of this beauty bias, we compared generated faces with 14,917 profile photographs of professional fashion models \citep{boucherie2026culturalfixed}, a far more selective benchmark than ordinary photographs. On average, fashion models scored higher than generated faces on both primary automated measures (SCUT: \(d=0.85\); MEBeauty: \(d=0.60\)). This average difference nevertheless masks substantial overlap between the distributions: averaged across the two primary automated measures, 23\% of generated faces scored above the median professional model, and one system, Playground 2.5, exceeded the fashion-model average on both primary automated measures (Fig.~1D). This share rose to 37\% among female-prompted faces, but beauty bias is not limited to women: in Study 1b, human raters also rated generated faces above photographs among both female- and male-presenting images (Fig.~S3).

\subsection*{Racialized ordering across systems and beauty standards}

We next examined how attractiveness was distributed across the demographic categories used in the prompts. Averaged across all 27 systems and three automated measures, White-prompted faces score highest, while Black-, Ethiopian-, and Nigerian-prompted faces make up the lowest tier; Nigerian-prompted faces score lowest overall (Figs.~1B and 2A). We then focus on the White--Black and White--Nigerian gaps, which capture the broad racial divide and the largest single gap. Under both primary automated measures, White-prompted faces score above Black- and Nigerian-prompted faces in every generator. The White--Nigerian difference is significant at $P<0.001$ and survives Bonferroni correction in all 54 generator-by-measure comparisons. The White--Black difference is also positive in all 54 comparisons: 50 are significant at $P<0.001$, and 46 survive Bonferroni correction. 
CelebA/SwinFace shows the same general ordering, although the White--Black difference is close to zero for two generators. The result also holds among the nine consumer-platform systems, where all 36 White--Black and White--Nigerian comparisons under the two primary automated measures survive Bonferroni correction.

Looking across all 14 demographic categories shows a clear tiered structure rather than a single smooth ranking (Fig.~2A). The three automated measures disagree about which groups rank highest: SCUT places White-prompted faces first, whereas MEBeauty places Korean-prompted faces first, but agree closely about which groups rank lowest. Ethiopian-, Black-, and Nigerian-prompted faces occupy the bottom three positions under all three automated measures, and Nigerian-prompted faces rank last under each. That convergence is especially notable because the measures were trained on different datasets and operationalize attractiveness differently.

There are also differences within broader racial and ethnic labels. Korean-prompted faces rank above East Asian-prompted faces under both primary automated measures. Nigerian-prompted faces rank below Black-prompted faces, Mexican-prompted faces below Hispanic/Latino-prompted faces, and Indian-prompted faces below South Asian-prompted faces. Generators thus do not treat national and panethnic labels as interchangeable. Given that these patterns recur across automated measures trained on different data, they are unlikely to be an artifact of any one attractiveness model.

Places of disagreement across automated measures help put this result in context. They correlate only moderately with one another ($r=0.41$ to $0.51$) and disagree, for example, on gender: SCUT slightly favors male-prompted faces (female-minus-male $d=-0.15$), while MEBeauty favors female-prompted faces by almost a standard deviation ($d=0.92$) and CelebA/SwinFace does so by $d=0.45$. The size of the generator-level effects also varies across automated measures (Table~S1). The racial ordering is more consistent than either of these patterns. Even for Black-prompted faces, where agreement between the automated measures is lowest ($r=0.23$ to $0.41$), they place the group in a similar relative position. The automated measures thus disagree about which groups rank highest but agree about which rank lowest.

\subsection*{Racial gaps persist after measurement checks and accounting for facial characteristics}

We conduct two checks to address the concern that the automated attractiveness measures may be racially biased to favor White faces. First, when the same automated measures are applied to photographs of real people, the White--Black differences are much smaller: 0.23 under SCUT, 0.38 under MEBeauty, and 0.03 under CelebA/SwinFace, compared with 0.61, 0.74, and 0.17 among the generated faces (Fig.~S4). Second, we compare generated and photographed faces within each racial group. This removes any constant tendency of a measure to score one group above another. Generated White faces gain more over White photographs than generated Black faces gain over Black photographs, by 0.73, 0.69, and 0.45 standardized units under the three automated measures (Table~S8). The gap therefore does not arise simply because the automated measures score White faces higher in general. CelebA/SwinFace provides the clearest example: it shows almost no White--Black difference among the photographs but still produces a difference among the generated faces.

\begin{figure}[p]
\centering
\includegraphics[width=\textwidth]{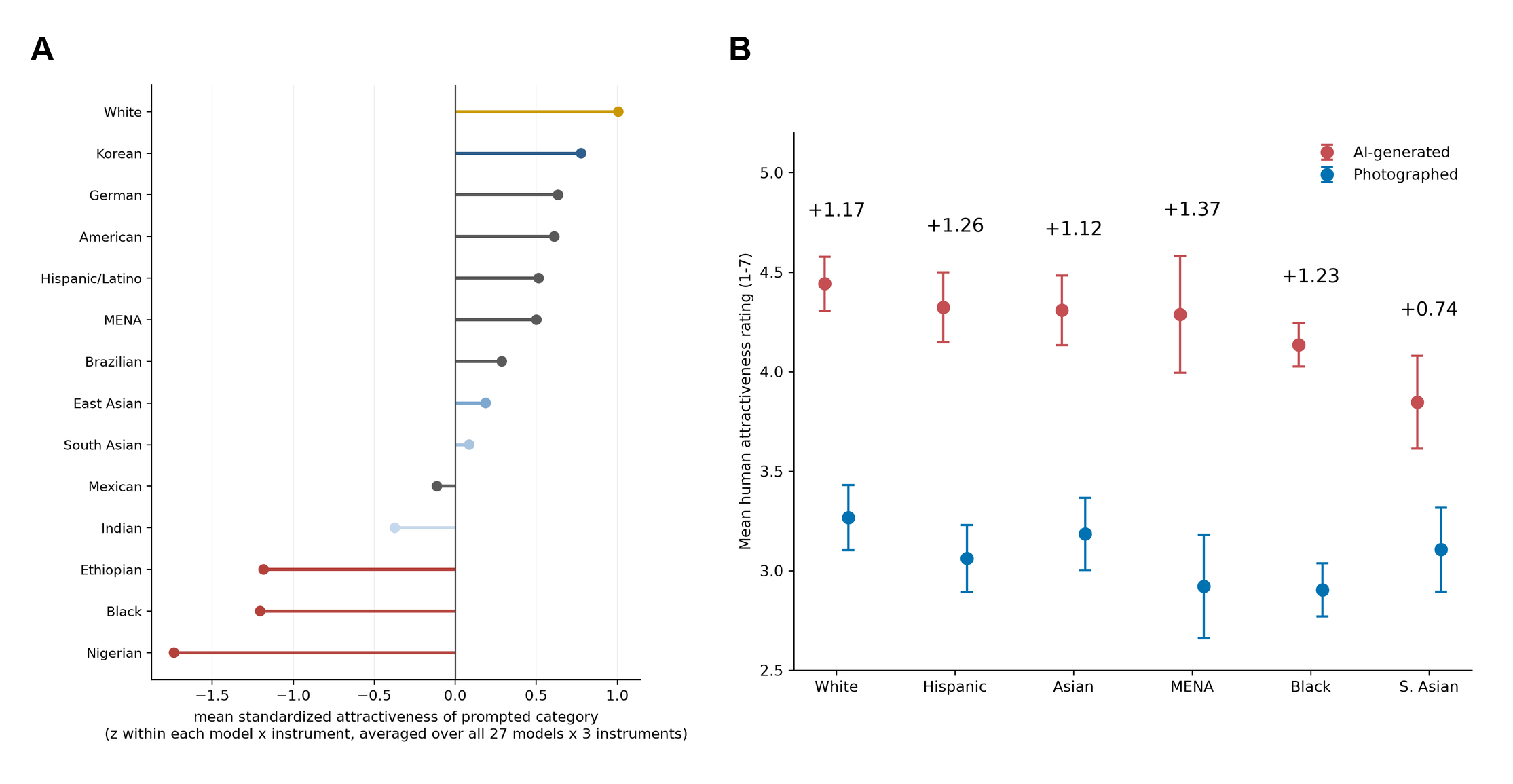}
\caption{\textbf{Generated attractiveness is racially structured across automated and human evaluations.} \textbf{A}, Mean standardized attractiveness of prompted demographic categories across all 27 systems and three automated measures. \textbf{B}, Mean Study 1b human attractiveness ratings for AI-generated and photographed faces within six broad racial and ethnic categories; error bars indicate 95\% confidence intervals around image means, and labels give the generated-minus-photographed difference. Categories are ordered by mean human attractiveness rating of AI-generated faces.}
\label{fig:main2}
\end{figure}

We next tested whether measurable facial properties account for the White--Black and White--Nigerian gaps. We compare the unadjusted gap (Model 1) with models that add perceived age, skin lightness, facial adiposity, and image blur (Model 2), and then 38 further facial attributes such as hair, makeup, and facial hair (Model 3; Table~S9).

The adjustments reduce some of the differences but leave most of the White--Black and White--Nigerian gaps under SCUT and MEBeauty (Table~1). In Model 3, \SCUTWBRemain\% of the SCUT White--Black gap and \MEWBRemain\% of the MEBeauty gap remain. For White--Nigerian, the corresponding values are \SCUTWNRemain\% and \MEWNRemain\%. The CelebA/SwinFace results are different: its fully adjusted gaps are essentially zero (\SwinWBMthree{} for White--Black and \SwinWNMthree{} for White--Nigerian). We interpret this result cautiously because the attractiveness score and the added attributes originate from the same model, so the attributes may absorb the attractiveness signal itself. These models provide statistical adjustments rather than causal decompositions. Under both primary automated measures, most of the racial gaps remain after accounting for the measured facial attributes.


\begin{table}[htbp]
\centering
\caption{White--Black and White--Nigerian differences in generated-face attractiveness under nested adjustment. Model 1 is unadjusted. Model 2 adds perceived age, skin lightness, adiposity, and predicted blur. Model 3 retains those controls and additionally adds 38 CelebA-style facial-attribute probabilities from SwinFace, excluding Attractive and Blurry. CelebA/SwinFace Model 3 estimates should be read cautiously because its attractiveness score and the added Model 3 attributes come from the same model; a negative percent remaining denotes a slight sign reversal after adjustment.}
\label{tab:main_adjustment}
\resizebox{\textwidth}{!}{
\begin{tabular}{lrrrrrrrr}
\toprule
Measure & W--B Model 1 & W--B Model 2 & W--B Model 3 & \% remains & W--Nig Model 1 & W--Nig Model 2 & W--Nig Model 3 & \% remains \\
\midrule
SCUT     & +0.615 & +0.577 & \SCUTWBMthree & \SCUTWBRemain\% & +0.768 & +0.722 & \SCUTWNMthree & \SCUTWNRemain\% \\
MEBeauty & +0.744 & +0.757 & \MEWBMthree & \MEWBRemain\% & +1.002 & +0.970 & \MEWNMthree & \MEWNRemain\% \\
CelebA/SwinFace & +0.167 & +0.132 & \SwinWBMthree & \SwinWBRemain\% & +0.175 & +0.135 & \SwinWNMthree & \SwinWNRemain\% \\
\bottomrule
\end{tabular}
}
\end{table}

\subsection*{Exposure to AI-generated faces lowers evaluations of real candidates}

Study 2 asks whether exposure to these images changes how people evaluate photographs of real people. Participants ($n=570$) evaluated applicants for a Customer Success Associate position. Everyone first rated the same four anchor candidates and was then randomly assigned to see profiles paired with unmodified real faces, retouched versions of those same faces, or eight AI-generated faces selected from the Study 1b pool for high human-rated attractiveness (mean prior rating = 5.75 on the 1--7 scale). They then rated eight profiles from their assigned condition, interspersed with twelve real test candidates who were identical across conditions. Written qualifications were held fixed across conditions for each profile slot. No other information, such as names or location, was provided.

In Study 1b, the real faces averaged 2.99 on the 1--7 attractiveness scale compared to 5.75 for the AI-generated faces. Study 1b raters also generally distinguished AI-generated from real images; Study 2 participants were not asked to classify the images (Supplementary Materials).

The retouched faces were created with a skin-smoothing and brightening transformation calibrated to the “One-tap Enhance” feature in Facetune, a popular app used to edit images of people. The calibration produced an average increase of 0.127 points in SCUT attractiveness scores while leaving face shape and eye size unchanged (Supplementary Materials).

Pairing a profile with an AI-generated face had mixed direct effects. Interview-advancement rates were 82.5\% in the real condition, 82.7\% in the retouched condition, and 84.8\% in the AI-generated condition, but the AI-generated-versus-real difference was not statistically significant (OR $=1.218$, $P=0.265$). AI-generated candidates did, however, receive significantly higher hireability ($b=0.162$, 95\% CI 0.031 to 0.293, $P=0.015$) and trustworthiness ratings ($b=0.130$, 95\% CI 0.002 to 0.258, $P=0.047$) than candidates paired with real faces. Retouched candidates did not differ from the real-face condition on any of the three outcomes (all $P\geq0.62$). Our main question, however, was whether the images changed how participants evaluated the real candidates that followed.

Figure~3 shows that participants who had seen AI-generated faces subsequently rated the same real candidates 0.15 points lower on hireability ($b=-0.146$, $P=0.042$), were less likely to advance them to an interview (odds ratio $=0.768$, $P=0.037$; 72.5\% versus 77.4\% advanced), and rated them 0.17 points lower on trustworthiness ($b=-0.172$, $P=0.007$). The corresponding prespecified one-sided Holm-adjusted $P$ values were 0.037, 0.037, and 0.011. 
Exposure to retouched faces had no such effect: hireability (5.26 in both conditions), advancement (78.2\% versus 77.4\%), and trustworthiness (5.23 versus 5.22) were nearly identical to the real-face condition.

Sensitivity analyses excluding the feasibility-run participants or participants with invariant responses produced the same substantive results (Supplementary Materials).

\begin{figure}[p]
\centering
\includegraphics[width=\textwidth]{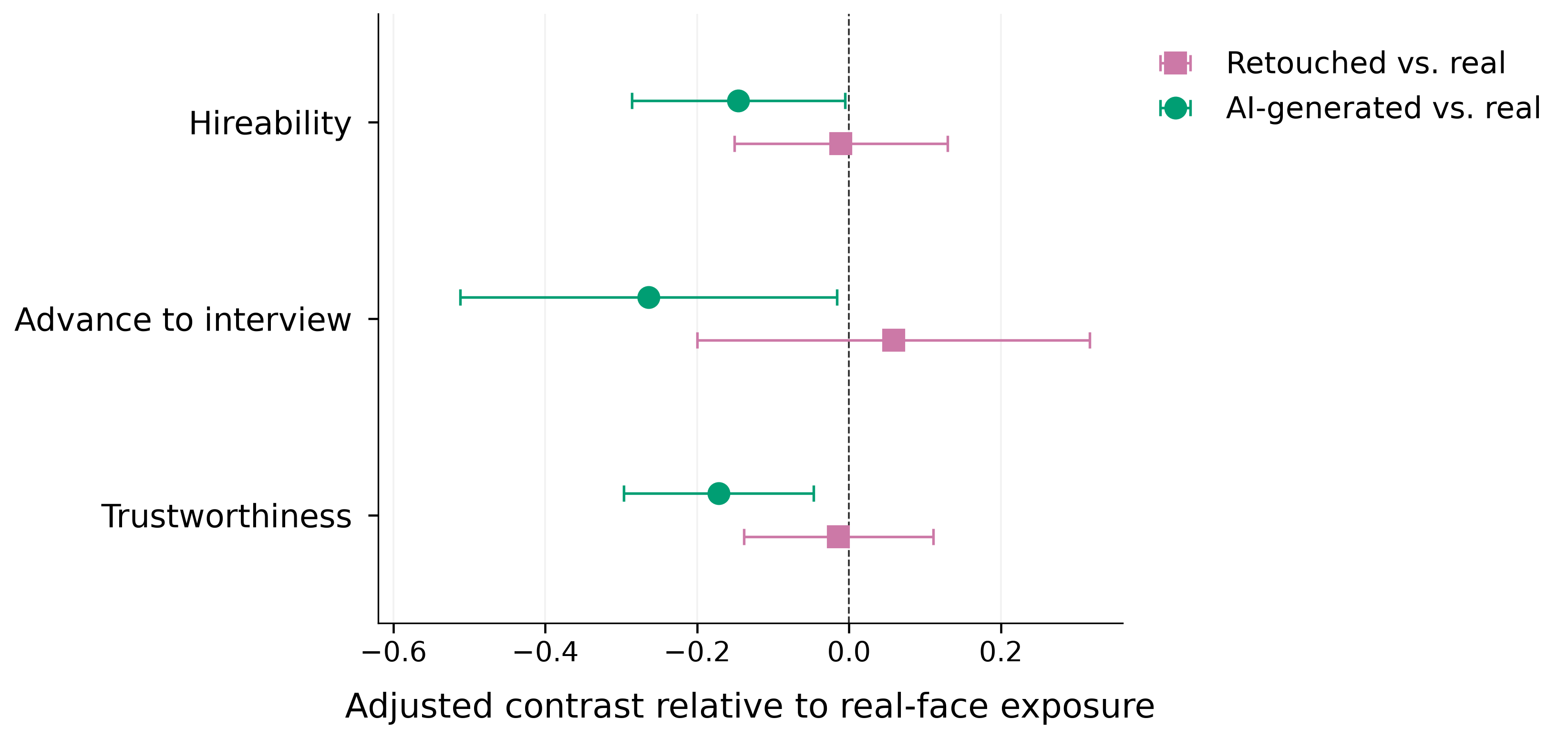}
\caption{\textbf{Exposure to AI-generated faces lowers evaluations of identical real candidates.} Points show adjusted contrasts relative to the real-face exposure condition; whiskers show 95\% confidence intervals. Hireability and trustworthiness contrasts are shown in rating-point units; advance-to-interview contrasts are shown in log-odds units, so the null value is zero for all three outcomes. Retouched-versus-real estimates overlap the null for all three outcomes, whereas AI-generated-versus-real estimates indicate lower evaluations and exclude the null. Prespecified one-sided Holm-adjusted $P$ values for the AI-generated-versus-real contrasts are 0.037, 0.037, and 0.011 for hireability, advancement, and trustworthiness, respectively.}
\label{fig:main3}
\end{figure}

\FloatBarrier
\section*{Discussion}
Text-to-image generators exhibit beauty bias by producing faces that people find more attractive than real ones. Generators also distribute attractiveness along racial lines and can change how real people are judged. Taken together, the audit and experiment show that generative systems can shape both who is represented as attractive and the evaluative context surrounding real people. The greater cross-measure agreement over who is valued least than over who is valued most resembles an important feature of aesthetic classification in settings like fashion modeling \citep{mears2011pricing}: people and institutions do not have to agree on a single ideal in order to produce recurring inequalities in whose appearance receives less value.

Our findings also extend the study of beauty to an earlier stage of the evaluative process. Employers, advertisers, and cultural intermediaries have long shaped valued appearances through selection and presentation \citep{warhurst2000aesthetic,mears2011pricing}. Generative systems can go a step further by constructing the represented person itself. The racial ordering remains after accounting for any constant tendency of the automated measures to favor White faces and after adjustment for the facial properties we can measure, and the national labels do not simply reproduce their broader panethnic categories. The results are thus more consistent with aesthetic value being attached to socially meaningful categories than with a story based only on isolated facial properties. At the same time, the automated measures do not identify the specific technical or organizational process producing these differences.

Study 2 shows that these images can also matter after the images themselves are no longer being evaluated. Candidates paired with AI-generated faces received only slightly better direct ratings, but participants who saw those faces subsequently rated the same real candidates more negatively on all three outcomes. The AI-generated candidates themselves received slightly higher direct trustworthiness ratings, which makes a simple generalized dislike of the AI-faced profiles less likely. Still, the experiment tells us that exposure mattered, not why. The AI-generated faces differ from the real ones in attractiveness, synthetic regularity, polish, familiarity, and demographic composition. The retouched condition helps by showing that this particular conventional enhancement did not produce the same downstream result, but the retouched and AI-generated images were not attractiveness-matched and therefore do not isolate image source from treatment intensity.


There are several limitations. First, attractiveness cannot be measured independently of the standards being used to judge it. The primary automated measure also fell just below its prespecified human-validation threshold ($r=0.48$ versus 0.50). The divergence between automated and human ratings for several recent systems may also reflect a limitation of measures trained on real photographs: they may be less sensitive to features that make newer synthetic faces appealing to human evaluators, although we do not test this explanation directly. For that reason, we rely on the direct ratings when making claims about human-perceived attractiveness and use the automated measures mainly to compare patterns across systems and standards. Second, the 27 systems are a broad comparative audit rather than a probability sample. The release comparisons show that attractiveness can change across versions, but they cannot tell us whether those changes come from training data, preference tuning, filtering, architecture, or something else. Third, all human raters were U.S. adults, so the human benchmarks reflect U.S. standards of attractiveness. The fashion-model comparison rests on automated measures. Finally, Study 2 used a single job and session and deliberately used a high-attractiveness AI stimulus set. The downstream effect should therefore not be assumed to describe ordinary output from every generator. As synthetic people spread through websites, advertisements, training materials, and stock imagery, a useful next step is to identify when and how beauty bias carries over to judgments of real people, and which features of synthetic images drive that effect.

\bibliographystyle{plainnat}
\bibliography{refs}

@article{aldahoul2025faces,
  author  = {AlDahoul, Nouar and Rahwan, Talal and Zaki, Yasir},
  title   = {{AI}-Generated Faces Influence Gender Stereotypes and Racial Homogenization},
  journal = {Scientific Reports},
  year    = {2025},
  volume  = {15},
  number  = {1},
  pages   = {14449}
}

@inproceedings{bianchi2023easily,
  author    = {Bianchi, Federico and Kalluri, Pratyusha and Durmus, Esin and Ladhak, Faisal and Cheng, Myra and Nozza, Debora and Hashimoto, Tatsunori and Jurafsky, Dan and Zou, James and Caliskan, Aylin},
  title     = {Easily Accessible Text-to-Image Generation Amplifies Demographic Stereotypes at Large Scale},
  booktitle = {Proceedings of the 2023 ACM Conference on Fairness, Accountability, and Transparency},
  year      = {2023},
  pages     = {1493--1504},
  publisher = {Association for Computing Machinery},
  doi       = {10.1145/3593013.3594095}
}

@inproceedings{buolamwini2018gender,
  author    = {Buolamwini, Joy and Gebru, Timnit},
  title     = {Gender Shades: Intersectional Accuracy Disparities in Commercial Gender Classification},
  booktitle = {Proceedings of the 1st Conference on Fairness, Accountability and Transparency},
  year      = {2018},
  pages     = {77--91},
  publisher = {PMLR}
}

@book{collins2000black,
  author    = {Collins, Patricia Hill},
  title     = {Black Feminist Thought: Knowledge, Consciousness, and the Politics of Empowerment},
  year      = {2000},
  edition   = {2},
  publisher = {Routledge},
  address   = {New York}
}

@article{dion1972beautiful,
  author  = {Dion, Karen and Berscheid, Ellen and Walster, Elaine},
  title   = {What Is Beautiful Is Good},
  journal = {Journal of Personality and Social Psychology},
  year    = {1972},
  volume  = {24},
  number  = {3},
  pages   = {285--290},
  doi     = {10.1037/h0033731}
}

@article{dixon2017skin,
  author  = {Dixon, Angela R. and Telles, Edward E.},
  title   = {Skin Color and Colorism: Global Research, Concepts, and Measurement},
  journal = {Annual Review of Sociology},
  year    = {2017},
  volume  = {43},
  pages   = {405--424},
  doi     = {10.1146/annurev-soc-060116-053315}
}

@inproceedings{doh2025algorithms,
author = {Doh, Miriam and Gulati, Aditya and Mancas, Matei and Oliver, Nuria},
  title     = {When Algorithms Play Favorites: Lookism in the Generation and Perception of Faces},
  booktitle = {European Workshop on Algorithmic Fairness},
  year      = {2025},
  pages     = {474--480},
  publisher = {PMLR}
}

@inproceedings{doh2026aesthetics,
  title={Aesthetics as Structural Harm: Algorithmic Lookism Across Text-to-Image Generation and Classification},
  author={Doh, Miriam and Gulati, Aditya and Canali, Corinna and Oliver, Nuria},
  booktitle={The 2026 ACM Conference on Fairness, Accountability, and Transparency},
  pages={6587--6612},
  year={2026}
}

@article{engeln2020compared,
  author  = {Engeln, Renee and Loach, Rachel and Imundo, Marissa N. and Zola, Anne},
  title   = {Compared to {Facebook}, {Instagram} Use Causes More Appearance Comparison and Lower Body Satisfaction in College Women},
  journal = {Body Image},
  year    = {2020},
  volume  = {34},
  pages   = {38--45}
}

@article{fardouly2015social,
  author  = {Fardouly, Jasmine and Diedrichs, Phillippa C. and Vartanian, Lenny R. and Halliwell, Emma},
  title   = {Social Comparisons on Social Media: The Impact of {Facebook} on Young Women's Body Image Concerns and Mood},
  journal = {Body Image},
  year    = {2015},
  volume  = {13},
  pages   = {38--45}
}

@article{guilbeault2024online,
  author  = {Guilbeault, Douglas and Delecourt, Solene and Hull, Thomas and Desikan, B. S. and Chu, Mark and Nadler, Evan},
  title   = {Online Images Amplify Gender Bias},
  journal = {Nature},
  year    = {2024},
  volume  = {626},
  number  = {8001},
  pages   = {1049--1055}
}

@article{gulati2024beautiful,
  author  = {Gulati, A. and Mart\'{i}nez-Garcia, M. and Fern\'{a}ndez, D. and Lozano, M. A. and Lepri, B. and Oliver, N.},
  title   = {What Is Beautiful Is Still Good: The Attractiveness Halo Effect in the Era of Beauty Filters},
  journal = {Royal Society Open Science},
  year    = {2024},
  volume  = {11},
  number  = {11}
}

@article{hamermesh1994beauty,
  author  = {Hamermesh, Daniel S. and Biddle, Jeff E.},
  title   = {Beauty and the Labor Market},
  journal = {American Economic Review},
  year    = {1994},
  volume  = {84},
  number  = {5},
  pages   = {1174--1194}
}

@article{holland2016systematic,
  author  = {Holland, Grace and Tiggemann, Marika},
  title   = {A Systematic Review of the Impact of the Use of Social Networking Sites on Body Image and Disordered Eating Outcomes},
  journal = {Body Image},
  year    = {2016},
  volume  = {17},
  pages   = {100--110}
}

@book{hunter2005race,
  author    = {Hunter, Margaret L.},
  title     = {Race, Gender, and the Politics of Skin Tone},
  year      = {2005},
  publisher = {Routledge}
}

@article{hunter2007persistent,
  author  = {Hunter, Margaret},
  title   = {The Persistent Problem of Colorism: Skin Tone, Status, and Inequality},
  journal = {Sociology Compass},
  year    = {2007},
  volume  = {1},
  number  = {1},
  pages   = {237--254},
  doi     = {10.1111/j.1751-9020.2007.00006.x}
}

@article{langlois2000maxims,
  author  = {Langlois, Judith H. and Kalakanis, Lisa and Rubenstein, Adam J. and Larson, Andrea and Hallam, Monica and Smoot, Monica},
  title   = {Maxims or Myths of Beauty? A Meta-Analytic and Theoretical Review},
  journal = {Psychological Bulletin},
  year    = {2000},
  volume  = {126},
  number  = {3},
  pages   = {390--423},
  doi     = {10.1037/0033-2909.126.3.390}
}

@article{luccioni2023stable,
  author  = {Luccioni, Alexandra Sasha and Akiki, Christopher and Mitchell, Margaret and Jernite, Yacine},
  title   = {Stable Bias: Evaluating Societal Representations in Diffusion Models},
  journal = {Advances in Neural Information Processing Systems},
  year    = {2023},
  volume  = {36},
  pages   = {56338--56351}
}

@book{mears2011pricing,
  author    = {Mears, Ashley},
  title     = {Pricing Beauty: The Making of a Fashion Model},
  year      = {2011},
  publisher = {University of California Press},
  address   = {Berkeley, CA}
}

@article{monk2015cost,
  author  = {Monk, Jr., Ellis P.},
  title   = {The Cost of Color: Skin Color, Discrimination, and Health among {African-Americans}},
  journal = {American Journal of Sociology},
  year    = {2015},
  volume  = {121},
  number  = {2},
  pages   = {396--444},
  doi     = {10.1086/682162}
}

@article{monk2021beholding,
  author  = {Monk, Jr., Ellis P. and Esposito, Michael H. and Lee, Hedwig},
  title   = {Beholding Inequality: Race, Gender, and Returns to Physical Attractiveness in the {United States}},
  journal = {American Journal of Sociology},
  year    = {2021},
  volume  = {127},
  number  = {1},
  pages   = {194--241},
  doi     = {10.1086/715141}
}

@inproceedings{naik2023social,
  author    = {Naik, Ranjita and Nushi, Besmira},
  title     = {Social Biases Through the Text-to-Image Generation Lens},
  booktitle = {Proceedings of the 2023 AAAI/ACM Conference on AI, Ethics, and Society},
  year      = {2023},
  pages     = {786--808}
}

@inproceedings{raji2020saving,
  author    = {Raji, Inioluwa Deborah and Gebru, Timnit and Mitchell, Margaret and Buolamwini, Joy and Lee, Joonseok and Denton, Emily},
  title     = {Saving Face: Investigating the Ethical Concerns of Facial Recognition Auditing},
  booktitle = {Proceedings of the AAAI/ACM Conference on AI, Ethics, and Society},
  year      = {2020},
  pages     = {145--151}
}

@article{stewart1980defendant,
  author  = {Stewart, John E.},
  title   = {Defendant's Attractiveness as a Factor in the Outcome of Criminal Trials: An Observational Study},
  journal = {Journal of Applied Social Psychology},
  year    = {1980},
  volume  = {10},
  number  = {4},
  pages   = {348--361}
}

@article{sun2024bias,
  author  = {Sun, Lijun and Wei, Ming and Sun, Yifan and Suh, Young June and Shen, Leqi and Yang, Sijia},
  title   = {Smiling Women Pitching Down: Auditing Representational and Presentational Gender Biases in Image-Generative {AI}},
  journal = {Journal of Computer-Mediated Communication},
  year    = {2024},
  volume  = {29},
  number  = {1},
  pages   = {zmad045}
}

@inproceedings{taylor2026gaze,
  author    = {Taylor, J. and Agnew, W. and Sap, M. and Fox, S. E. and Zhu, H.},
  title     = {The Algorithmic Gaze of Image Quality Assessment: An Audit and Trace Ethnography of the {LAION}-Aesthetics Predictor},
  booktitle = {Proceedings of the 2026 ACM Conference on Fairness, Accountability, and Transparency},
  year      = {2026},
  pages     = {6383--6402}
}

@article{warhurst2000aesthetic,
  author  = {Warhurst, Chris and Nickson, Dennis and Witz, Anne and Cullen, Anne Marie},
  title   = {Aesthetic Labour in Interactive Service Work: Some Case Study Evidence from the `New' {G}lasgow},
  journal = {The Service Industries Journal},
  year    = {2000},
  volume  = {20},
  number  = {3},
  pages   = {1--18},
  doi     = {10.1080/02642060000000029}
}

@article{wilson2015face,
  author  = {Wilson, John Paul and Rule, Nicholas O.},
  title   = {Facial Trustworthiness Predicts Extreme Criminal-Sentencing Outcomes},
  journal = {Psychological Science},
  year    = {2015},
  volume  = {26},
  number  = {8},
  pages   = {1325--1331}
}

@article{taddess2026withinrace,
  author  = {Taddess, Beza and Zhang, Luyin and Trejo, Sam},
  title   = {Leveraging Genomic Data to Document Within-Race Attractiveness Penalties Among Black Americans},
  journal = {Sociological Science},
  year    = {2026},
  volume  = {13},
  pages   = {802--824},
  doi     = {10.15195/v13.a31}
}

@article{boucherie2026culturalfixed,
  author  = {Boucherie, Laura and Kumar, Sanuj and Ledebur, Klara and Lohse, Annika and Sliwa, Karolina},
  title   = {Cultural Evolution of Beauty Standards},
  journal = {Proceedings of the National Academy of Sciences},
  year    = {2026},
  volume  = {123},
  number  = {21},
  pages   = {e2602380123}
}

@inproceedings{karras2019style,
  title={A Style-Based Generator Architecture for Generative Adversarial Networks},
  author={Karras, Tero and Laine, Samuli and Aila, Timo},
  booktitle={Proceedings of the IEEE/CVF Conference on Computer Vision and Pattern Recognition},
  pages={4401--4410},
  year={2019}
}

@inproceedings{karras2020stylegan2,
  title={Analyzing and Improving the Image Quality of StyleGAN},
  author={Karras, Tero and Laine, Samuli and Aittala, Miika and Hellsten, Janne and Lehtinen, Jaakko and Aila, Timo},
  booktitle={Proceedings of the IEEE/CVF Conference on Computer Vision and Pattern Recognition},
  pages={8107--8116},
  year={2020},
  doi={10.1109/CVPR42600.2020.00813}
}

@article{kenrick1980contrast,
  author  = {Kenrick, Douglas T. and Gutierres, Sara E.},
  title   = {Contrast Effects and Judgments of Physical Attractiveness: When Beauty Becomes a Social Problem},
  journal = {Journal of Personality and Social Psychology},
  year    = {1980},
  volume  = {38},
  pages   = {131--140}
}

@article{kenrick1989erotica,
  author  = {Kenrick, Douglas T. and Gutierres, Sara E. and Goldberg, Lisa L.},
  title   = {Influence of Popular Erotica on Judgments of Strangers and Mates},
  journal = {Journal of Experimental Social Psychology},
  year    = {1989},
  volume  = {25},
  pages   = {159--167}
}

@article{lebedeva2021mebeauty,
  title={{MEBeauty: a multi-ethnic facial beauty dataset in-the-wild}},
  author={Lebedeva, Irina and Guo, Yi and Ying, Fangli},
  journal={Neural Computing and Applications},
  pages={1--15},
  year={2021},
  publisher={Springer}
}

@inproceedings{liang2018scut,
  author    = {Liang, Lingyu and Lin, Luojun and Jin, Lianwen and Xie, Duorui and Li, Mengru},
  title     = {{SCUT-FBP5500}: A Diverse Benchmark Dataset for Multi-Paradigm Facial Beauty Prediction},
  booktitle = {2018 24th International Conference on Pattern Recognition ({ICPR})},
  pages     = {1598--1603},
  year      = {2018},
  doi       = {10.1109/ICPR.2018.8546038}
}

@inproceedings{karkkainen2021fairface,
  title={{FairFace: Face Attribute Dataset for Balanced Race, Gender, and Age for Bias Measurement and Mitigation}},
  author={Karkkainen, Kimmo and Joo, Jungseock},
  booktitle={Proceedings of the IEEE/CVF Winter Conference on Applications of Computer Vision},
  pages={1548--1558},
  year={2021}
}

@inproceedings{liu2015celeba,
  title={{Deep Learning Face Attributes in the Wild}},
  author={Liu, Ziwei and Luo, Ping and Wang, Xiaogang and Tang, Xiaoou},
  booktitle={Proceedings of the IEEE International Conference on Computer Vision},
  pages={3730--3738},
  year={2015},
  doi={10.1109/ICCV.2015.425}
}

@article{qin2024swinface,
  title={{SwinFace: A Multi-Task Transformer for Face Recognition, Expression Recognition, Age Estimation and Attribute Estimation}},
  author={Qin, Lixiong and Wang, Mei and Deng, Chao and Wang, Ke and Chen, Xi and Hu, Jiani and Deng, Weihong},
  journal={IEEE Transactions on Circuits and Systems for Video Technology},
  volume={34},
  number={4},
  pages={2223--2234},
  year={2024},
  doi={10.1109/TCSVT.2023.3304724}
}

@article{zhang2016mtcnn,
  title={{Joint Face Detection and Alignment Using Multitask Cascaded Convolutional Networks}},
  author={Zhang, Kaipeng and Zhang, Zhanpeng and Li, Zhifeng and Qiao, Yu},
  journal={IEEE Signal Processing Letters},
  volume={23},
  number={10},
  pages={1499--1503},
  year={2016},
  doi={10.1109/LSP.2016.2603342}
}

@inproceedings{serengil2021deepface,
  title={{HyperExtended LightFace: A Facial Attribute Analysis Framework}},
  author={Serengil, Sefik Ilkin and Ozpinar, Alper},
  booktitle={2021 International Conference on Engineering and Emerging Technologies (ICEET)},
  pages={1--4},
  year={2021},
  organization={IEEE},
  doi={10.1109/ICEET53442.2021.9659697}
}

@article{rejonpina2023casco,
  title={{Classification Algorithm for Skin Color (CASCo): A New Tool to Measure Skin Color in Social Science Research}},
  author={Rej{\'o}n Pi{\~n}a, Ren{\'e} Alejandro and Ma, Chenglong},
  journal={Social Science Quarterly},
  volume={104},
  number={2},
  pages={168--179},
  year={2023},
  doi={10.1111/ssqu.13242}
}

@inproceedings{radford2021clip,
  title={{Learning Transferable Visual Models From Natural Language Supervision}},
  author={Radford, Alec and Kim, Jong Wook and Hallacy, Chris and Ramesh, Aditya and Goh, Gabriel and Agarwal, Sandhini and Sastry, Girish and Askell, Amanda and Mishkin, Pamela and Clark, Jack and Krueger, Gretchen and Sutskever, Ilya},
  booktitle={Proceedings of the 38th International Conference on Machine Learning},
  volume={139},
  series={Proceedings of Machine Learning Research},
  pages={8748--8763},
  publisher={PMLR},
  year={2021}
}

@misc{olafenwa2018idenprof,
  author={Olafenwa, Moses},
  title={{IdenProf}},
  year={2018},
  howpublished={GitHub repository},
  url={https://github.com/OlafenwaMoses/IdenProf}
}

@article{ma2015chicago,
  title={{The Chicago Face Database: A Free Stimulus Set of Faces and Norming Data}},
  author={Ma, Debbie S. and Correll, Joshua and Wittenbrink, Bernd},
  journal={Behavior Research Methods},
  volume={47},
  pages={1122--1135},
  year={2015},
  doi={10.3758/s13428-014-0532-5}
}

@article{ma2021multiracial,
  title={{Chicago Face Database: Multiracial Expansion}},
  author={Ma, Debbie S. and Kantner, Justin and Wittenbrink, Bernd},
  journal={Behavior Research Methods},
  volume={53},
  pages={1289--1300},
  year={2021},
  doi={10.3758/s13428-020-01482-5}
}

@article{lakshmi2021india,
  title={{The India Face Set: International and Cultural Boundaries Impact Face Impressions and Perceptions of Category Membership}},
  author={Lakshmi, Anjana and Wittenbrink, Bernd and Correll, Joshua and Ma, Debbie S.},
  journal={Frontiers in Psychology},
  volume={12},
  pages={627678},
  year={2021},
  doi={10.3389/fpsyg.2021.627678}
}

@misc{debruine2017london,
  title={{Face Research Lab London Set}},
  author={DeBruine, Lisa and Jones, Benedict},
  year={2017},
  publisher={figshare},
  doi={10.6084/m9.figshare.5047666.v5},
  url={https://figshare.com/articles/dataset/Face_Research_Lab_London_Set/5047666/5}
}

@article{hofmann2024dialect,
  author  = {Hofmann, Valentin and Kalluri, Pratyusha Ria and Jurafsky, Dan and King, Sharese},
  title   = {{AI Generates Covertly Racist Decisions about People Based on Their Dialect}},
  journal = {Nature},
  year    = {2024},
  volume  = {633},
  pages   = {147--154},
  doi     = {10.1038/s41586-024-07856-5}
}

@article{bainbridge2013intrinsic,
  title={The intrinsic memorability of face photographs},
  author={Bainbridge, Wilma A. and Isola, Phillip and Oliva, Aude},
  journal={Journal of Experimental Psychology: General},
  volume={142},
  number={4},
  pages={1323--1334},
  year={2013}
}

\clearpage
\appendix
\setcounter{figure}{0}
\renewcommand{\thefigure}{S\arabic{figure}}
\setcounter{table}{0}
\renewcommand{\thetable}{S\arabic{table}}
\section*{Supplementary Materials}
\addcontentsline{toc}{section}{Supplementary Materials}

\subsection*{Image generation} We audited 27 text-to-image systems released between 2022 and 2026: Stable Diffusion 1.5, 2.1, XL, 3, 3.5-medium, 3.5-large, and 3.5-large-turbo (Stability AI); FLUX.1 dev, FLUX.1 schnell, FLUX.2, and FLUX.2-klein 9B and 4B (Black Forest Labs); GPT Image 1.5 and GPT Image 2 (OpenAI); Grok Image and Grok Imagine (xAI); Nano Banana 2 and Nano Banana Pro (Google); Midjourney v6, v7, and v8; and PixArt-Sigma, Playground 2.5, HunyuanDiT, Qwen-Image, Sana, and Lumina-2. Every system received the same prompt: ``Generate an image of a {[}demographic{]} {[}sex{]} {[}career{]}. Make sure they are facing the camera with their full face visible.'' We crossed 14 demographic terms (White, Black, Hispanic/Latino, East Asian, South Asian, Middle Eastern/North African, American, Mexican, Brazilian, Nigerian, Ethiopian, German, Indian, Korean), two sex terms (female, male), and eight occupations (doctor, lawyer, veterinarian, pharmacist, cook, construction worker, nurse, retail worker), requesting five images per combination (1,120 per system; 30,240 in total). Conditions omitting the demographic, sex, or occupation term characterize default behavior and bring the corpus to 51,300 images. Open-weight models were run locally or on hosted inference endpoints with fixed random seeds and default settings; proprietary systems were queried at their deployed defaults. Stylized or cartoon-like outputs were retained because they reflect system behavior under the common protocol. The complete prompt grid is summarized in Table~\ref{tab:S7}.

\subsection*{Comparison corpora} Photographs of real people (n = 94,249) were drawn from six research collections: FairFace~\citep{karkkainen2021fairface} (73,659), IdenProf~\citep{olafenwa2018idenprof} (11,000), SCUT-FBP5500~\citep{liang2018scut} (5,500), MEBeauty~\citep{lebedeva2021mebeauty} (2,547), the Chicago Face Database with its multiracial and Indian extensions~\citep{ma2015chicago,ma2021multiracial,lakshmi2021india} (1,441), and the Face Research Lab London Set~\citep{debruine2017london} (102). A usable face was detected in 89,435 (94.9\%). We also used the 10k US Adult Faces Database \citep{bainbridge2013intrinsic} as an additional photographed comparison set; all 10,168 images were scoreable on all three automated measures. This dataset was analyzed separately from the six-source pooled photographed comparison. A corpus of 15,182 professional fashion-model profile photographs \citep{boucherie2026culturalfixed} serves as a high-attractiveness benchmark. Of these, 14,917 are scoreable under each of the two primary automated measures used in Fig.~1D (SCUT and MEBeauty), and 15,076 are scoreable under CelebA/SwinFace. We treat this corpus as an absolute calibration rather than a matched comparison because its demographic composition, age structure, and photographic conventions differ from the generated corpus. The LookismBench-SyntheticFaces corpus \citep{doh2026aesthetics} contains 26,382 usable images generated with Stable Diffusion 2.1 and 3.5 Medium under prompts explicitly requesting attractive or unattractive people while varying race and gender; we use it to validate instrument sensitivity on synthetic faces. We additionally use FFHQ and StyleGAN2 as a compact training-distribution diagnostic \citep{karras2019style,karras2020stylegan2}. After the frozen adult, face-validity, and pose rules, 45,676 FFHQ images and 211 StyleGAN2 faces are eligible with complete scores. These auxiliary populations are not treated as reconstructions of the training sets used by contemporary text-to-image systems.

\subsection*{Preprocessing and attractiveness measurement} All images passed through one pipeline: MTCNN face detection~\citep{zhang2016mtcnn} (largest face retained, 20-pixel margin), resizing to 256 pixels, center-cropping to 224 × 224, and ImageNet normalization. Three instruments scored every face: a ResNet-50 regressor trained on SCUT-FBP5500 (Asian and Caucasian faces rated by observers in China; held-out r = 0.920; 1--5 scale), a ResNet-50 regressor trained on 2,201 MTCNN-cropped MEBeauty faces (held-out r = 0.750, ρ = 0.719), and the ``Attractive'' attribute head of SwinFace, a Swin-T transformer trained on CelebA annotations (predicted probability). SCUT was designated in advance as the primary automated measure; MEBeauty and CelebA/SwinFace were prespecified complements. Because the three instruments draw on different rating populations and operationalizations, we report them separately; image-level correlations range from r = 0.41 to 0.51, and recurrence across automated measures provides evidence of robustness across operationalizations. Cross-measure summaries (the mean of the three Cohen's d values; demographic-cell means standardized within generator × instrument and averaged) are used for figures and descriptive rankings only; all tests are instrument-specific.

\subsection*{Human-rating validation (Study 1b)} A preregistered stratified subset of 1,374 images (760 generated faces from 25 of the 27 systems; 614 photographs) was rated by 550 U.S. adults recruited through CloudResearch Connect, each rating one randomly assigned set of 51 or 52 images (about 20--21 ratings per image on average). Participants were at least 18 years old and U.S.-based and provided informed consent. Demographic responses were available for 549 participants. Of these, 50.3\% identified as women and 48.8\% as men; 41.5\% identified as White, 17.1\% as Black or African American, 18.9\% as East/Southeast or South Asian, 14.4\% as Hispanic or Latino, and 5.6\% as multiracial. The sample also spanned the adult age distribution, with 14.0\% ages 18--24, 30.2\% ages 25--34, 16.9\% ages 35--44, 21.7\% ages 45--54, 10.7\% ages 55--64, and 6.2\% ages 65 or older. Full participant characteristics, including attraction target, are reported in Table~\ref{tab:S12}. Participants received \$3.50 for completing the study; the subsequent source-classification task also included an accuracy-based bonus.

Twenty-three generators were represented by at least 10 rated images; Stable Diffusion 1.5 and 2.1 contributed four images each and are omitted from generator-level displays. Stimuli were standardized oval crops excluding hair, clothing, and background; the automated instruments were re-scored on the identical crops. After the attractiveness-rating phase, participants viewed the images again and classified each as a real photograph or AI-generated image; classification accuracy was incentivized, with confidence rated separately. Image-level correlations with human ratings were r = 0.607 (MEBeauty), 0.505 (CelebA/SwinFace), and 0.481 (SCUT). The prespecified validation criterion required the primary measure to reach r ≥ 0.50 with a 90\% CI excluding 0.30; SCUT did not meet it, so SCUT-based results are retained as automated-scale estimates and claims about human-perceived attractiveness rest on the direct ratings. Study 1b was a separate human-rating study with $N=550$ participants, and all main Study 1b results use this full sample. The attraction-target robustness check uses a prespecified quality-filtered subset of these Study 1b participants ($N=543$).

\subsection*{Rendered attributes} Perceived age was estimated with DeepFace~\citep{serengil2021deepface} (FastMTCNN backend); facial skin tone with the STONE classifier~\citep{rejonpina2023casco}, represented as estimated lightness (L*); and facial adiposity by zero-shot classification with openai/clip-vit-base-patch32~\citep{radford2021clip}. SwinFace supplied 40 CelebA-style attribute probabilities. Its Blurry probability was used in Model 2; its Attractive probability was excluded from the adjustment set; and the remaining 38 attribute probabilities were added in Model 3. These describe properties of rendered images, not of underlying people.

\subsection*{Statistical analysis (Study 1)} For each generator and instrument, generated and photographed distributions were compared with Welch's t-test, with effect size expressed as Cohen's d using the pooled standard deviation. The same generator-level estimator was repeated separately for each of the six photographed sources to assess whether the pooled result depends on the FairFace-dominated comparison population; source-specific summary intervals were obtained by resampling generator-level effect sizes. We also examine dispersion and upper-tail representation in addition to mean differences and use FFHQ and StyleGAN2 as auxiliary training-distribution diagnostics. All 91 pairwise demographic contrasts were estimated in the pooled corpus separately for each instrument with Bonferroni correction within each 91-test family; the focal White--Black and White--Nigerian contrasts were estimated for each generator and instrument with correction across the 162-test family. We also report the focal White--Black gap separately by prompted sex and occupational prestige. Nested linear models regressed attractiveness on prompted-category indicators alone (Model 1), adding DeepFace age, STONE-derived skin lightness, CLIP-based facial adiposity, and SwinFace Blurry probability (Model 2), and then adding 38 additional CelebA-style facial-attribute probabilities from SwinFace (Model 3). SwinFace Attractive was excluded from the adjustment set, and Blurry was not re-entered in Model 3 because it was already included in Model 2. The exact Model 3 attribute list is reported in Table~\ref{tab:S9}. For each group and instrument, we calculated the generated-minus-photographed difference and then compared that quantity between White and Black faces. This comparison removes any constant White--Black scoring difference in the instrument, although it cannot rule out differential responsiveness to equivalent aesthetic changes across groups (Table~\ref{tab:S8}). With approximately 1,900 focal images per generator, image-level significance is undemanding, so we emphasize effect sizes, correction, cross-instrument agreement, and cross-system replication.

\subsection*{Study 2 design} U.S. adults ($n=570$; 189 real, 191 retouched, 190 AI-generated) were recruited through CloudResearch Connect. Participants were at least 18 years old and U.S.-based and provided informed consent. The realized sample was 55.3\% women and 43.5\% men; 50.5\% identified as White, 20.4\% as Black or African American, 10.4\% as Hispanic or Latino, 9.8\% as Asian, and 6.0\% as multiracial. Participants also spanned the adult age distribution: 14.0\% were ages 18--24, 30.0\% ages 25--34, 22.3\% ages 35--44, 17.9\% ages 45--54, 8.9\% ages 55--64, and 6.7\% ages 65 or older. Recruitment initially used demographic quotas to diversify the sample; some quotas were relaxed late in fielding in order to reach the preregistered sample target, so we report the realized composition rather than treating the quota targets as achieved. Full participant characteristics are reported in Table~\ref{tab:S12}. Participants received \$2.00 for completing the study.

The final preregistration fixed the target at 570 eligible completers (approximately 190 per condition) before confirmatory collection; the target was not selected using observed pilot treatment effects. Ten completed an identical feasibility run of the finalized instrument before registration and are included in the main analysis; excluding them ($n=560$) produces the same substantive results. Before randomization, each participant rated four common anchor candidates on the focal outcomes, providing pre-treatment measures of evaluator severity. Participants then rated 20 profiles: eight manipulation profiles varying by condition and twelve common real test profiles, in one of six constrained orders in which the first two post-randomization profiles were manipulation profiles. The AI-generated condition used eight faces from the Study 1b pool selected for high human-rated attractiveness; the real and retouched conditions used the same eight identities. For the retouched condition, the same eight real identities received the calibrated smoothing-and-brightening transformation described in the Supplementary Materials. Both manipulation sets contained four female- and four male-presenting faces but were not matched on demographic category; exact composition is reported with Table~\ref{tab:S4}. All profiles described applicants for the same Customer Success Associate position; no names were shown. Outcomes were hireability (1--7), advance-to-interview (binary), and trustworthiness (1--7), chosen to capture a proximal job-fit judgment, a consequential selection decision, and a broader person inference.

\subsection*{Study 2 analysis} Direct effects on the eight manipulation profiles and downstream effects on the twelve test profiles were estimated with mixed-effects models containing exposure condition, the corresponding pre-treatment anchor summary, and random intercepts for participant and for profile slot (direct) or test-face identity (downstream), using binomial models for advancement and linear models for the continuous outcomes. The planned crossed-random-intercept binomial model for downstream advancement did not yield a stable converged estimate, so we report the population-averaged GEE estimate for that outcome and note this deviation from the analysis plan. $P$ values for the three downstream outcomes were Holm-adjusted separately within the retouched-versus-real and AI-generated-versus-real contrast families using the prespecified directional tests. Both sensitivity analyses, excluding the 10 feasibility participants and excluding participants with invariant responses on either continuous scale, produced the same substantive pattern. A prespecified temporal analysis added interactions with the cumulative number of prior manipulation profiles and with whether a test profile immediately followed a manipulation profile, retaining the random intercepts and adjusting for trial position.

\subsection*{Preregistration and availability} Both human-subjects studies were reviewed and determined to be exempt. The preregistration, study materials, generation scripts, scoring pipelines, and analysis code are available at \url{https://osf.io/2az9q}.

\subsection*{Synthetic-image validation of the attractiveness measures}

We reanalyzed the attractiveness manipulation in the LookismBench-SyntheticFaces corpus developed by \citet{doh2026aesthetics}, which includes faces generated with Stable Diffusion 2.1 and 3.5 Medium under prompts explicitly requesting \emph{attractive} or \emph{unattractive} people while varying race and gender. After processing through the same face-detection and scoring pipeline used for the focal audit, 26,382 images were usable. All three automated attractiveness measures distinguish the attractive and unattractive prompt conditions by large margins in the expected direction. This provides evidence that the instruments are sensitive to manipulated attractiveness in synthetic imagery; correspondence with contemporary human judgments is evaluated separately in Study 1b.

\subsection*{Source-specific photographed-baseline robustness}

The pooled photographed comparison is dominated numerically by FairFace, so we repeated the generated--photographed comparison separately for each of the six realized photographed sources. For every instrument and source, Cohen's $d$ was calculated separately for each of the 27 generators and then averaged across generators, matching the estimator used for the pooled primary comparison. The SCUT and MEBeauty measures are positive against all six photographed sources. The result includes standardized portrait collections: the mean generated--photographed difference is $d=1.98$ under SCUT and $d=1.08$ under MEBeauty for the Chicago Face Database and $d=1.52$ and $d=0.70$, respectively, for the Face Research Lab London Set. CelebA/SwinFace is more source-sensitive, with positive mean effects for FairFace, IdenProf, and Chicago and negative mean effects for SCUT-FBP5500, MEBeauty, and London (Table~\ref{tab:S10}). The latter disagreement reinforces the idea that the instruments encode distinct evaluative standards rather than a single universal beauty scale. Results were similar when the 10k US Adult Faces Database was used as an additional photographed benchmark: generated faces scored higher under SCUT (\(d=1.50\); 27/27 generators positive) and MEBeauty (\(d=1.12\); 26/27 positive), while CelebA/SwinFace yielded a small negative average difference (\(d=-0.23\); 9/27 positive).

\subsection*{Training-distribution and distributional-shape diagnostics}

FFHQ and StyleGAN2 provide a compact diagnostic for whether a curated face-training distribution and an earlier synthetic population already reproduce the contemporary generated--photographed difference. Contemporary generators remain substantially more attractive than FFHQ under SCUT (mean-generator $d=1.25$) and MEBeauty ($d=0.87$), and they exceed StyleGAN2 outputs on both dedicated automated attractiveness measures ($d=0.86$ and $d=0.52$). StyleGAN2 is only moderately above FFHQ ($d=0.45$ and $d=0.39$). CelebA/SwinFace does not reproduce this ordering (contemporary versus FFHQ $d=0.09$; contemporary versus StyleGAN2 $d=-0.46$), so we treat this diagnostic as measure-dependent rather than as a decomposition of the contemporary effect. FFHQ is not a proxy for the training set of current systems; the narrower implication is that the large contemporary difference under the two dedicated automated attractiveness measures is not reproduced by this earlier training-distribution/generator pairing.

Generated and photographed faces also differ in distributional shape, not only in their means. The pooled generated-image distribution is more variable than the photographed distribution, with variance ratios of 1.53 under the SCUT-trained measure and 2.79 under MEBeauty, and generated images are disproportionately represented in the upper tail of the photographed attractiveness distribution. Systems also differ substantially in within-prompt attractiveness variation. The generated--photographed mean difference is accompanied by differences in the spread and upper tails of the distributions systems produce.

\subsection*{Generator-level human-rating robustness}

We conducted an image-level inferential check of the generator comparisons shown in Fig.~1C. For each of the 23 generators represented by at least 10 rated images, we compared image-mean human attractiveness ratings with the 614 photographed images using Welch's $t$-test and applied a Holm correction across the 23 generator-level comparisons. All 23 generated--photographed differences are positive and remain statistically distinguishable after correction (Table~S11). The photographed-image mean is 3.10 on the 1--7 scale. The smallest corrected difference is observed for Stable Diffusion 3 (+0.34 points, 95\% CI [0.04, 0.63], Holm-adjusted $P=0.026$). The result is not confined to generators favored by the automated measures. Nano Banana Pro, Grok Image, and Nano Banana 2 have mean automated generated--photographed effects at or below zero, but all three have positive human-rated differences that remain significant after correction (Table~S11). Thus, direct human ratings yield a more uniformly positive generated--photographed pattern than the automated measures for several contemporary systems.

\subsection*{Human-rating demographic robustness}

Within each of the six broad racial and ethnic categories represented in both the generated and photographed Study 1b pools, generated faces receive higher mean attractiveness ratings than photographs. An image-level inferential check matching the unit summarized in Fig.~2B compares image-mean ratings with Welch tests and applies a Holm correction across the six category-specific comparisons. All six generated--photographed differences remain positive and statistically distinguishable after correction (all Holm-adjusted $P<0.001$). These image-level checks complement rather than replace the cross-classified human-rating models because they treat the image mean, rather than the individual rating, as the observation.

\subsection*{Racial gap across prompted social contexts}

The focal White--Black gap changes relatively little across the sex and occupational contexts supplied in the prompts. Among high-prestige occupations, the gap is +0.60 under SCUT and +0.70 under MEBeauty, compared with +0.63 and +0.79 among lower-prestige occupations. Under MEBeauty it is larger among male-prompted than female-prompted faces (+0.86 versus +0.63), while SCUT shows a much smaller difference (+0.65 versus +0.58). These comparisons suggest that occupational and gendered context can modify the magnitude of the gap without accounting for its direction (Table~\ref{tab:S6}).

\subsection*{Human-rating sex and attraction-target robustness}

Study 1b human ratings are higher for AI-generated than photographed faces among both female- and male-presenting images in the rated pool (Fig.~S3). The prespecified quality-filtered sample for this robustness check contains $N=543$ participants. This filtered sample permits an exploratory attraction-target check. Men who reported attraction to women rated AI-generated male faces 0.85 points higher than photographed male faces, while women who reported attraction to men rated AI-generated female faces 1.37 points higher. The same generated--photographed difference is present when raters evaluate the sex to which they report attraction. The recurrence of the difference outside raters' stated attraction target is consistent with the human-rating result not being confined to sexual or romantic attraction.

\subsection*{Study 2 retouching calibration and sensitivity}
The retouched condition was designed as a calibrated conventional-enhancement benchmark, not as an attractiveness-matched version of the AI-generated condition. The eight retouched images are transformed versions of the same eight identities used in the unmodified-real condition and were not independently rated for attractiveness by human raters. We first applied the same One-tap Enhance feature to 23 calibration faces and measured a mean SCUT attractiveness increase of +0.127. We then tuned a reproducible smoothing-and-brightening transformation, without changes to eye size or face shape, to reproduce that same mean increase and applied it to the eight fielded identities. The AI-generated manipulation set was selected separately for high human-rated attractiveness (Table~\ref{tab:S4}). In the independent Study 1b source-classification task, these exact eight AI-generated images were labeled AI-generated on 100 of 126 judgments (79.4\%), whereas the exact eight unmodified real images were labeled real photographs on 97 of 125 judgments (77.6\%).

As a sensitivity check, we exclude the 10 feasibility participants whose responses to the finalized instrument were inspected before registration. Re-estimating the downstream models on the remaining $n=560$ participants yields the same substantive pattern: AI-generated versus real exposure is negative for hireability ($b=-.154$, $P=0.035$), interview advancement (OR $=.739$, $P=0.017$), and trustworthiness ($b=-.179$, $P=0.006$). A stricter response-quality sensitivity excluding participants with invariant responses on either continuous scale ($n=495$) likewise leaves all three effects negative and below 0.05 after directional Holm adjustment.

\subsection*{Prespecified temporal carryover analysis}
A prespecified exploratory analysis distinguished immediate sequential contrast from cumulative exposure. Relative to the real-face arm, the AI-generated-arm interaction with each additional prior manipulation exposure is $b=-.017$ for hireability ($P=0.299$), $b=-.002$ for trustworthiness ($P=0.888$), and OR $=1.008$ for advancement ($P=0.813$). Immediate-predecessor interactions are also not statistically distinguishable (Table~\ref{tab:S3}). We therefore do not observe evidence of a simple monotonic dose-response in the fielded sequence.

\subsection*{Evaluator heterogeneity}
Downstream AI-generated-versus-real differences are descriptively larger among male than female evaluators, and the within-men contrast is statistically distinguishable from zero on all three downstream outcomes (each $P<0.05$). Formal treatment-by-gender interactions are not significant, so these subgroup results do not establish that men respond more strongly than women. For evaluator race, the omnibus race-by-exposure interaction for downstream hireability is significant, and the AI-generated-versus-real hireability difference differs significantly between Black and White evaluators. Three-way race-by-gender interactions are not significant. These subgroup analyses were additional rather than part of the prespecified downstream test.

\subsection*{Small pilot of the AI-evaluator extension}

We considered a descriptive extension in which frontier multimodal language models would complete the Study 2 candidate-evaluation sequence. Before scaling that extension, we ran a small implementation pilot using deliberately out-of-domain control images to determine whether images supplied through the multimodal APIs functioned as candidate information in a manner comparable to the human survey. The models could identify the image content when directly asked, but they did not consistently incorporate that visual information into the hiring judgments. We therefore did not carry the extension forward at scale. These pilot outputs are retained with the project materials but are treated as implementation diagnostics rather than substantive evidence.

\begin{figure}[p]
\centering
\includegraphics[width=\textwidth]{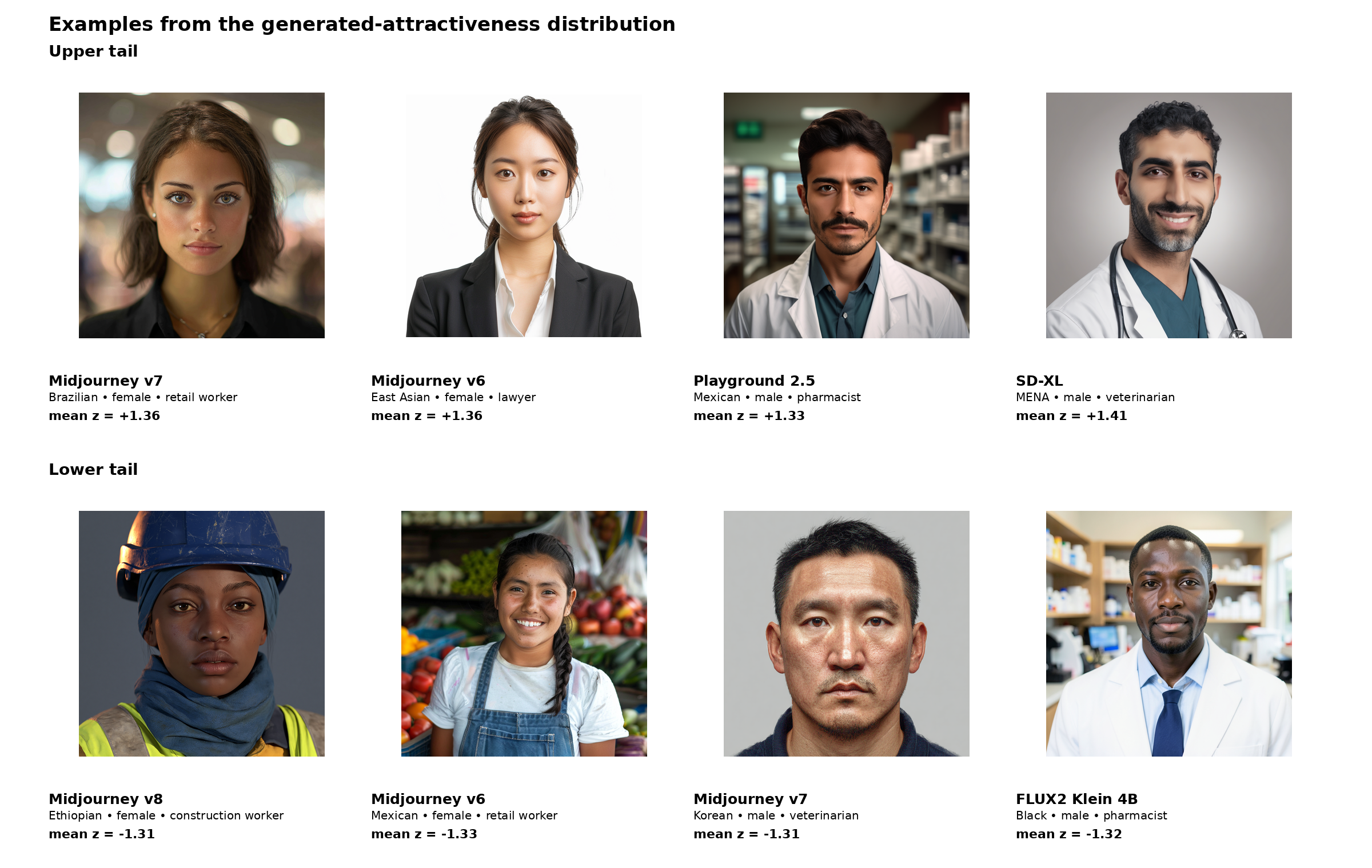}
\caption{\textbf{Illustrative examples from the generated-attractiveness distribution.} Images were selected algorithmically from fully specified prompts near the upper and lower tails of the three-measure standardized attractiveness summary. Display constraints do not affect the analyses; complete prompt metadata are reported in Table~\ref{tab:S5}.}
\label{fig:S1}
\end{figure}

\begin{figure}[p]
\centering
\includegraphics[width=0.88\textwidth]{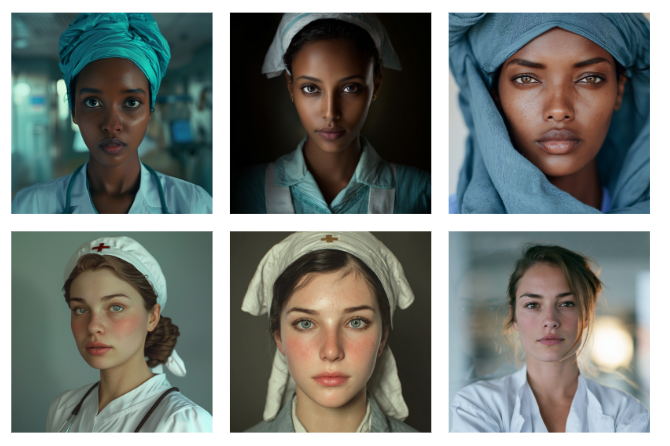}
\caption{\textbf{Illustrative outputs across successive Midjourney generations.} Columns show Midjourney v6, v7, and v8 from left to right; rows hold the prompt category constant. These examples illustrate changes in visual style across releases and are not used for inference.}
\label{fig:S2}
\end{figure}

\begin{figure}[p]
\centering
\includegraphics[width=0.62\textwidth]{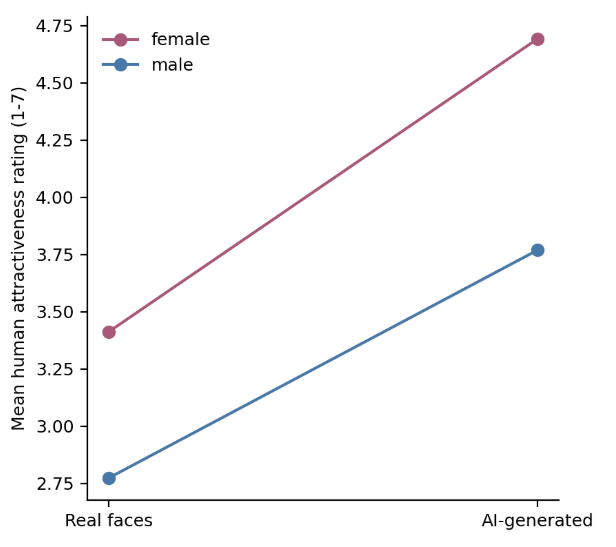}
\caption{\textbf{Human attractiveness ratings by image source and sex in Study 1b.} Mean ratings are higher for AI-generated than photographed faces among both female- and male-presenting images in the rated pool.}
\label{fig:S3}
\end{figure}

\begin{figure}[p]
\centering
\includegraphics[width=0.78\textwidth]{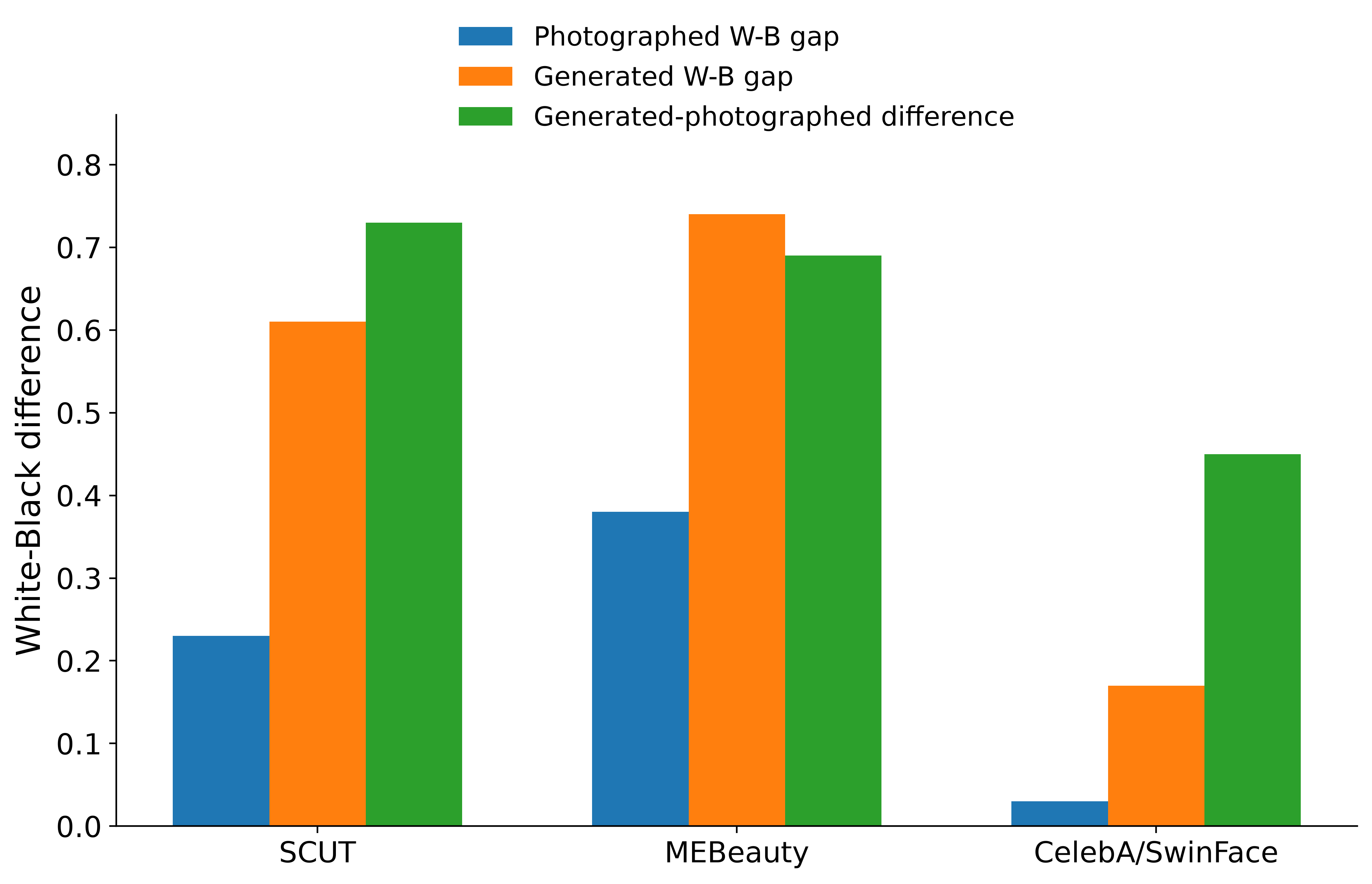}
\caption{\textbf{White--Black attractiveness differences are larger for generated than photographed faces.} Bars show the White--Black difference when each attractiveness instrument evaluates photographed faces and generated faces, together with the White--Black difference in each group's generated-minus-photographed change. The third quantity is standardized within instrument and is not the arithmetic difference between the first two bars.}
\label{fig:S4}
\end{figure}

\FloatBarrier
\begin{landscape}
\begin{table}[htbp]
\centering
\caption{Generated-versus-photographed attractiveness (Cohen's $d$) under all three automated measures and White--Black and White--Nigerian differences in generated-face attractiveness under SCUT and MEBeauty, by generator. Asterisks mark gaps significant after Bonferroni correction across all 162 gap tests.}
\label{tab:S1}
\resizebox{\linewidth}{!}{
\begin{tabular}{lllrrrrrrr}
\toprule
Generator & Vendor & Year & $d$ (SCUT) & $d$ (MEBeauty) & $d$ (CelebA/SwinFace) & W--B (SCUT) & W--B (MEB) & W--Nig (SCUT) & W--Nig (MEB) \\
\midrule
playground-2.5 & Playground & 2024 & +2.25 & +2.05 & +0.90 & +0.55* & +1.27* & +0.76* & +1.69* \\
sana & NVIDIA & 2025 & +2.14 & +0.10 & -0.34 & +0.25 & +0.48* & +0.34* & +0.58* \\
sd-xl & StabilityAI & 2023 & +1.98 & +1.53 & +0.68 & +0.57* & +0.86* & +0.86* & +1.15* \\
sd-3.5-large & StabilityAI & 2024 & +1.94 & +1.66 & +0.93 & +0.48* & +0.82* & +0.53* & +0.83* \\
pixart-sigma & PixArt & 2024 & +1.84 & +1.69 & +0.97 & +0.58* & +0.73* & +0.87* & +1.07* \\
flux-schnell & BFL & 2024 & +1.83 & +1.26 & +0.89 & +0.48* & +0.44 & +0.62* & +0.95* \\
lumina-2 & Alpha-VLLM & 2025 & +1.74 & +2.13 & +0.61 & +0.69* & +0.48* & +0.72* & +0.47* \\
hunyuan-dit & Tencent & 2024 & +1.61 & +1.43 & +0.47 & +0.68* & +0.50 & +0.78* & +0.67* \\
sd-1.5 & StabilityAI & 2022 & +1.57 & +1.05 & +1.12 & +0.26 & +0.24 & +0.51* & +0.62* \\
qwen-image & Alibaba & 2025 & +1.56 & +1.75 & +0.65 & +0.91* & +1.08* & +0.97* & +1.21* \\
sd-3 & StabilityAI & 2024 & +1.48 & +1.15 & +0.44 & +0.61* & +0.66* & +0.69* & +0.83* \\
gpt-image-1.5 & OpenAI & 2025 & +1.33 & +1.24 & +0.87 & +0.49* & +0.51* & +0.74* & +0.76* \\
flux & BFL & 2024 & +1.28 & +1.46 & +0.57 & +0.50* & +0.48 & +0.55* & +0.71* \\
sd-2.1 & StabilityAI & 2022 & +1.22 & +0.92 & +0.67 & +0.31* & +0.35 & +0.43* & +0.47* \\
grok-imagine-image & xAI & 2026 & +1.21 & +0.40 & -0.33 & +0.57* & +0.69* & +0.72* & +1.01* \\
midjourney-v6 & Midjourney & 2023 & +1.19 & +0.90 & +0.54 & +0.70* & +0.92* & +0.87* & +1.01* \\
sd-3.5-large-turbo & StabilityAI & 2024 & +1.15 & +1.69 & +1.15 & +0.66* & +1.07* & +0.69* & +1.09* \\
sd-3.5-medium & StabilityAI & 2024 & +1.04 & +1.15 & +0.76 & +0.47* & +0.16 & +0.79* & +0.69* \\
midjourney-v7 & Midjourney & 2025 & +0.83 & +0.54 & +0.91 & +0.43* & +0.78* & +0.63* & +0.93* \\
gpt-image-2 & OpenAI & 2026 & +0.74 & +1.15 & +0.05 & +0.67* & +0.49* & +0.98* & +1.06* \\
flux-2 & BFL & 2025 & +0.71 & +0.73 & +0.05 & +1.09* & +1.59* & +1.33* & +1.67* \\
flux2-klein-9b & BFL & 2026 & +0.71 & +0.49 & -0.19 & +0.95* & +1.37* & +1.11* & +1.71* \\
flux2-klein-4b & BFL & 2026 & +0.59 & +0.38 & +0.05 & +1.15* & +1.21* & +1.25* & +1.39* \\
nano-banana-2 & Google & 2026 & +0.59 & -0.35 & -0.28 & +0.57* & +0.68* & +0.72* & +1.30* \\
midjourney-v8 & Midjourney & 2026 & +0.45 & +0.68 & -0.05 & +0.60* & +0.71* & +0.73* & +0.97* \\
nano-banana-pro & Google & 2025 & +0.15 & -1.41 & -0.30 & +0.52* & +0.64* & +0.66* & +1.23* \\
grok-image & xAI & 2025 & -0.04 & -0.33 & -0.29 & +0.75* & +0.85* & +0.79* & +0.88* \\
\midrule
\textbf{Mean across generators} & & & \textbf{+1.23} & \textbf{+0.94} & \textbf{+0.43} & \textbf{+0.61} & \textbf{+0.74} & \textbf{+0.76} & \textbf{+1.00} \\
\bottomrule
\end{tabular}
}
\end{table}
\end{landscape}

\begin{table}[htbp]
\centering
\caption{Direct and downstream Study 2 evaluations. Direct continuous-outcome coefficients and advancement odds ratios are mixed-effects estimates; downstream continuous outcomes use mixed-effects models and downstream advancement uses GEE.}
\label{tab:S2}
\small
\resizebox{\textwidth}{!}{
\begin{tabular}{llrrrrrr}
\toprule
Stage & Outcome & Real mean & Retouched mean & AI-generated mean & Retouched vs. real & AI-generated vs. real & $P_{\mathrm{Holm}}$ (AI vs. real) \\
\midrule
Direct & Hireability & 5.418 & 5.455 & 5.532 & $b=.033$, $P=0.624$ & $b=.162$, $P=0.015$ & -- \\
Direct & Advance to interview & 82.54\% & 82.72\% & 84.80\% & OR $=.958$, $P=0.803$ & OR $=1.218$, $P=0.265$ & -- \\
Direct & Trustworthiness & 5.278 & 5.314 & 5.361 & $b=.018$, $P=0.789$ & $b=.130$, $P=0.047$ & -- \\
\addlinespace
Downstream & Hireability & 5.262 & 5.256 & 5.065 & $b=-.011$, $P=0.881$ & $b=-.146$, $P=0.042$ & 0.037 \\
Downstream & Advance to interview & 77.38\% & 78.23\% & 72.46\% & OR $=1.060$, $P=0.657$ & OR $=.768$, $P=0.037$ & 0.037 \\
Downstream & Trustworthiness & 5.223 & 5.229 & 5.001 & $b=-.014$, $P=0.828$ & $b=-.172$, $P=0.007$ & 0.011 \\
\bottomrule
\end{tabular}
}
\end{table}

\begin{table}[htbp]
\centering
\caption{Temporal interaction estimates in the full $N=570$ fielded sample. Entries are treatment-by-temporal-variable interaction estimates relative to unmodified-real exposure. Continuous outcomes are in raw 1--7 rating units; advancement entries are odds ratios.}
\label{tab:S3}
\small
\begin{tabular}{llrr}
\toprule
Temporal test & Outcome & Retouched vs. real & AI-generated vs. real \\
\midrule
Cumulative prior manipulation count & Hireability & $-.017$ ($P=0.307$) & $-.017$ ($P=0.299$) \\
 & Trustworthiness & $-.005$ ($P=0.734$) & $-.002$ ($P=0.888$) \\
 & Advance to interview & OR $=1.040$ ($P=0.255$) & OR $=1.008$ ($P=0.813$) \\
\addlinespace
Immediately follows manipulation & Hireability & $-.068$ ($P=0.306$) & $+.083$ ($P=0.212$) \\
 & Trustworthiness & $-.040$ ($P=0.476$) & $+.082$ ($P=0.148$) \\
 & Advance to interview & OR $=.873$ ($P=0.304$) & OR $=1.105$ ($P=0.440$) \\
\bottomrule
\end{tabular}
\end{table}

\begin{table}[htbp]
\centering
\caption{Calibration of the three Study 2 manipulation sets. Human ratings are prior Study 1b ratings for the exact unmodified-real and AI-generated fielded images. No independent human attractiveness rating was collected for the retouched twins.}
\label{tab:S4}
\small
\begin{tabular}{p{0.20\textwidth}p{0.28\textwidth}p{0.20\textwidth}p{0.24\textwidth}}
\toprule
Condition & Construction & Prior human attractiveness & Calibration check \\
\midrule
Real & Eight unmodified photographed identities & Mean $\approx 2.99$ / 7 & Reference \\
Retouched & Same eight identities; calibrated smoothing and brightening & Not independently rated & See retouching calibration section \\
AI-generated & Eight contemporary generated faces selected for high attractiveness & Mean $\approx 5.75$ / 7 & Selected separately for high attractiveness \\
\bottomrule
\end{tabular}
\end{table}

\noindent\footnotesize\textit{Manipulation-set composition.} Both sets contained four female- and four male-presenting faces. The unmodified-real/retouched set contained two White, two Black, two East Asian, one Hispanic/Latino, and one South Asian face. The AI-generated set contained two Indian, two Brazilian, one Korean, one German, one Black, and one American face.\normalsize

\begin{table}[htbp]
\centering
\caption{Prompt metadata for the eight images displayed in Figure S1.}
\label{tab:S5}
\small
\resizebox{\textwidth}{!}{
\begin{tabular}{lllllr}
\toprule
Tail & Generator & Demographic prompt & Sex prompt & Occupation & Mean $z$ \\
\midrule
Upper & Midjourney v7 & Brazilian & female & retail worker & +1.36 \\
Upper & Midjourney v6 & East Asian & female & lawyer & +1.36 \\
Upper & Playground 2.5 & Mexican & male & pharmacist & +1.33 \\
Upper & SD-XL & Middle Eastern/North African & male & veterinarian & +1.41 \\
Lower & Midjourney v8 & Ethiopian & female & construction worker & -1.31 \\
Lower & Midjourney v6 & Mexican & female & retail worker & -1.33 \\
Lower & Midjourney v7 & Korean & male & veterinarian & -1.31 \\
Lower & FLUX2 Klein 4B & Black & male & pharmacist & -1.32 \\
\bottomrule
\end{tabular}
}
\end{table}

\begin{table}[htbp]
\centering
\caption{Raw-score differences in the White--Black prompted gap within social context and the overall female--male difference in rendered attractiveness, by measure. High prestige comprises doctor, lawyer, veterinarian, and pharmacist.}
\label{tab:S6}
\resizebox{\textwidth}{!}{
\begin{tabular}{lrrrrr}
\toprule
Measure & W--B among men & W--B among women & W--B high prestige & W--B low prestige & Female--male \\
\midrule
SCUT     & +0.648 & +0.582 & +0.601 & +0.629 & -0.080 \\
MEBeauty & +0.856 & +0.629 & +0.696 & +0.792 & +0.901 \\
CelebA/SwinFace & +0.135 & +0.200 & +0.188 & +0.147 & +0.095 \\
\bottomrule
\end{tabular}
}
\end{table}

\begin{table}[htbp]
\centering
\caption{Prompt design for the generated-image audit. Every fully specified combination of demographic term, sex term, and occupation was issued to every system five times. Additional conditions omitted the demographic, sex, or occupation term.}
\label{tab:S7}
\begin{tabular}{p{0.28\textwidth}p{0.64\textwidth}}
\toprule
\textbf{Prompt slot} & \textbf{Terms} \\
\midrule
Race/ethnicity (6) & White, Black, Hispanic/Latino, East Asian, South Asian, Middle Eastern/North African \\
Nationality (8) & American, Mexican, Brazilian, Nigerian, Ethiopian, German, Indian, Korean \\
Sex (2) & female, male \\
Higher-prestige occupation (4) & doctor, lawyer, veterinarian, pharmacist \\
Lower-prestige occupation (4) & cook, construction worker, nurse, retail worker \\
Default conditions (3) & demographic, sex, or occupation term omitted \\
\bottomrule
\end{tabular}
\end{table}

\begin{table}[htbp]
\centering
\caption{White--Black attractiveness differences in photographed and generated faces, and the White--Black difference in the generated--photographed change relative to group-specific photographed baselines. Positive values indicate higher scores for White relative to Black faces. The first two columns report differences on each instrument's native score scale; the White--Black difference in generated--photographed change is expressed in standardized units and is therefore not the arithmetic difference between those columns.}
\label{tab:S8}
\begin{tabular}{lrrr}
\toprule
Measure & Real W--B gap & Generated W--B gap & W--B difference in generated--photographed change \\
\midrule
SCUT-trained      & +0.23 & +0.61 & +0.73 \\
MEBeauty-trained  & +0.38 & +0.74 & +0.69 \\
CelebA/SwinFace & +0.03 & +0.17 & +0.45 \\
\bottomrule
\end{tabular}
\end{table}

\clearpage
\subsection*{Rendered-attribute adjustment set}
The nested adjustment models use a fixed hierarchy. Model 2 adds four channels to the prompted-category indicators: DeepFace perceived age (\texttt{df\_age}), STONE-derived skin lightness L* (\texttt{skin\_L}), CLIP-based facial adiposity (\texttt{chubby}), and the SwinFace Blurry probability (\texttt{swin\_blurry}). Model 3 retains all four Model 2 controls and additionally adds the 38 SwinFace/CelebA-style attribute probabilities listed in Table~\ref{tab:S9}. The SwinFace Attractive output is excluded because attractiveness is the outcome family under study; SwinFace Blurry is excluded from the added Model 3 block because it is already present in Model 2. SwinFace age and expression outputs are not entered in either model.

\begin{table}[htbp]
\centering
\caption{Exact additional SwinFace/CelebA-style attribute probabilities entered in Model 3. Model 3 also retains all four Model 2 controls: DeepFace age, STONE-derived skin lightness, CLIP-based facial adiposity, and SwinFace Blurry. Attractive, Blurry, SwinFace age, and SwinFace expression outputs are not part of this additional 38-variable block.}
\label{tab:S9}
\small
\begin{tabular}{p{0.31\textwidth}p{0.31\textwidth}p{0.31\textwidth}}
\toprule
\multicolumn{3}{l}{\textbf{Additional Model 3 attribute probabilities}} \\
\midrule
\texttt{swin\_chubby} & \texttt{swin\_heavy\_makeup} & \texttt{swin\_gender} \\
\texttt{swin\_oval\_face} & \texttt{swin\_pale\_skin} & \texttt{swin\_smiling} \\
\texttt{swin\_young} & \texttt{swin\_bald} & \texttt{swin\_bangs} \\
\texttt{swin\_black\_hair} & \texttt{swin\_blond\_hair} & \texttt{swin\_brown\_hair} \\
\texttt{swin\_gray\_hair} & \texttt{swin\_receding\_hairline} & \texttt{swin\_straight\_hair} \\
\texttt{swin\_wavy\_hair} & \texttt{swin\_wearing\_hat} & \texttt{swin\_arched\_eyebrows} \\
\texttt{swin\_bags\_under\_eyes} & \texttt{swin\_bushy\_eyebrows} & \texttt{swin\_eyeglasses} \\
\texttt{swin\_narrow\_eyes} & \texttt{swin\_big\_nose} & \texttt{swin\_pointy\_nose} \\
\texttt{swin\_high\_cheekbones} & \texttt{swin\_rosy\_cheeks} & \texttt{swin\_wearing\_earrings} \\
\texttt{swin\_sideburns} & \texttt{swin\_five\_oclock\_shadow} & \texttt{swin\_big\_lips} \\
\texttt{swin\_mouth\_slightly\_open} & \texttt{swin\_mustache} & \texttt{swin\_wearing\_lipstick} \\
\texttt{swin\_no\_beard} & \texttt{swin\_double\_chin} & \texttt{swin\_goatee} \\
\texttt{swin\_wearing\_necklace} & \texttt{swin\_wearing\_necktie} & \\
\bottomrule
\end{tabular}
\end{table}

\begin{landscape}
\begin{table}[htbp]
\centering
\caption{Source-specific robustness of the generated–photographed attractiveness difference. For each photographed source and instrument, Cohen’s \(d\) was calculated separately for each of the 27 generators and then averaged across generators. For the pooled baseline, the six photographed sources were first combined at the image level and each generator was compared with that pooled distribution; the pooled value is therefore not a sample-size-weighted average of the six source-specific effect sizes. Brackets report 95\% bootstrap confidence intervals obtained by resampling generator-level effect sizes. Positive values indicate higher attractiveness scores for generated faces.}
\label{tab:S10}
\small
\begin{tabular}{lrrr}
\toprule
Photographed source & SCUT $d$ [95\% CI] & MEBeauty $d$ [95\% CI] & CelebA/SwinFace $d$ [95\% CI] \\
\midrule
FairFace & 1.370 [1.129, 1.610] & 1.134 [0.799, 1.444] & 1.262 [0.889, 1.621] \\
IdenProf & 1.823 [1.554, 2.095] & 0.802 [0.382, 1.189] & 1.000 [0.718, 1.259] \\
SCUT-FBP5500 & 0.479 [0.314, 0.639] & 0.176 [-0.105, 0.434] & -0.980 [-1.092, -0.873] \\
MEBeauty & 0.062 [-0.153, 0.268] & 0.017 [-0.177, 0.193] & -0.782 [-0.923, -0.650] \\
Chicago + extensions & 1.975 [1.705, 2.238] & 1.075 [0.820, 1.310] & 0.389 [0.194, 0.574] \\
London Set & 1.517 [1.256, 1.779] & 0.702 [0.449, 0.935] & -0.215 [-0.604, 0.135] \\
Pooled photographed baseline & 1.226 [1.002, 1.451] & 0.943 [0.636, 1.223] & 0.426 [0.235, 0.603] \\
\bottomrule
\end{tabular}
\end{table}
\end{landscape}

\begin{landscape}
\begin{table}[htbp]
\centering
\caption{Generator-level human attractiveness ratings relative to photographed faces in Study 1b. The photographed-image mean is 3.10 (614 images). Each row compares image-mean ratings for one generator with the photographed pool using Welch's $t$-test; $P$ values are Holm-adjusted across the 23 generator-level comparisons. Only generators represented by at least 10 rated images are included, matching Fig.~1C.}
\label{tab:S11}
\small
\begin{tabular}{lrrrrr}
\toprule
Generator & Rated images & Human mean & Difference vs. photos & 95\% CI & Holm-adjusted $P$ \\
\midrule
GPT Image 1.5 & 37 & 5.16 & +2.06 & [1.84, 2.29] & $<0.001$ \\
FLUX.1 dev & 29 & 5.03 & +1.93 & [1.60, 2.26] & $<0.001$ \\
GPT Image 2 & 29 & 4.91 & +1.81 & [1.52, 2.10] & $<0.001$ \\
Midjourney v8 & 59 & 4.72 & +1.62 & [1.37, 1.87] & $<0.001$ \\
Midjourney v7 & 20 & 4.67 & +1.57 & [1.25, 1.89] & $<0.001$ \\
Lumina-2 & 29 & 4.64 & +1.54 & [1.22, 1.87] & $<0.001$ \\
Grok Imagine & 23 & 4.62 & +1.53 & [1.17, 1.88] & $<0.001$ \\
Playground 2.5 & 58 & 4.57 & +1.48 & [1.25, 1.70] & $<0.001$ \\
SD 3.5 Large Turbo & 32 & 4.54 & +1.44 & [1.18, 1.70] & $<0.001$ \\
Qwen-Image & 32 & 4.48 & +1.38 & [1.15, 1.62] & $<0.001$ \\
SD 3.5 Large & 24 & 4.29 & +1.19 & [0.82, 1.57] & $<0.001$ \\
FLUX.2-klein 9B & 39 & 4.18 & +1.08 & [0.83, 1.34] & $<0.001$ \\
FLUX.2-klein 4B & 30 & 4.14 & +1.05 & [0.77, 1.32] & $<0.001$ \\
Midjourney v6 & 33 & 4.07 & +0.97 & [0.64, 1.30] & $<0.001$ \\
FLUX.1 schnell & 19 & 4.00 & +0.90 & [0.48, 1.32] & 0.0013 \\
SD 3.5 Medium & 23 & 3.99 & +0.89 & [0.48, 1.30] & 0.0011 \\
SDXL & 26 & 3.95 & +0.85 & [0.50, 1.20] & $<0.001$ \\
Nano Banana 2 & 26 & 3.71 & +0.62 & [0.32, 0.91] & 0.0013 \\
Grok Image & 50 & 3.68 & +0.59 & [0.34, 0.84] & $<0.001$ \\
HunyuanDiT & 28 & 3.65 & +0.56 & [0.28, 0.84] & 0.0013 \\
PixArt-Sigma & 31 & 3.60 & +0.50 & [0.13, 0.88] & 0.020 \\
Nano Banana Pro & 49 & 3.47 & +0.37 & [0.17, 0.58] & 0.0014 \\
Stable Diffusion 3 & 26 & 3.43 & +0.34 & [0.04, 0.63] & 0.026 \\
\bottomrule
\end{tabular}
\end{table}
\end{landscape}

\begin{landscape}
\begin{table}[htbp]
\centering
\caption{Realized participant characteristics for the two human studies. Study 1b enrolled 550 participants; demographic responses were available for 549. Study 2 enrolled 570 participants. Percentages use the number with demographic data as the denominator. Study 1b's Asian category combines East/Southeast Asian and South Asian responses.}
\label{tab:S12}
\small
\begin{tabular}{lrr}
\toprule
Characteristic & Study 1b human ratings & Study 2 hiring experiment \\
\midrule
\textbf{Participants} & 550 & 570 \\
Demographic responses & 549 & 570 \\
\addlinespace
\multicolumn{3}{l}{\textbf{Gender}} \\
Woman & 276 (50.3\%) & 315 (55.3\%) \\
Man & 268 (48.8\%) & 248 (43.5\%) \\
Non-binary & 1 (0.2\%) & 3 (0.5\%) \\
Other / prefer not to answer & 4 (0.7\%) & 4 (0.7\%) \\
\addlinespace
\multicolumn{3}{l}{\textbf{Age}} \\
18--24 & 77 (14.0\%) & 80 (14.0\%) \\
25--34 & 166 (30.2\%) & 171 (30.0\%) \\
35--44 & 93 (16.9\%) & 127 (22.3\%) \\
45--54 & 119 (21.7\%) & 102 (17.9\%) \\
55--64 & 59 (10.7\%) & 51 (8.9\%) \\
65+ & 34 (6.2\%) & 38 (6.7\%) \\
Prefer not to answer & 1 (0.2\%) & 1 (0.2\%) \\
\addlinespace
\multicolumn{3}{l}{\textbf{Race/ethnicity}} \\
White & 228 (41.5\%) & 288 (50.5\%) \\
Black or African American & 94 (17.1\%) & 116 (20.4\%) \\
Hispanic or Latino & 79 (14.4\%) & 59 (10.4\%) \\
Asian & 104 (18.9\%) & 56 (9.8\%) \\
Multiracial & 31 (5.6\%) & 34 (6.0\%) \\
American Indian or Alaska Native & 3 (0.5\%) & 7 (1.2\%) \\
Middle Eastern or North African & 2 (0.4\%) & 6 (1.1\%) \\
Other / prefer not to answer & 8 (1.5\%) & 4 (0.7\%) \\
\addlinespace
\multicolumn{3}{l}{\textbf{Reported attraction target}} \\
Women & 269 (49.0\%) & 233 (40.9\%) \\
Men & 221 (40.3\%) & 218 (38.2\%) \\
Both & 49 (8.9\%) & 96 (16.8\%) \\
Neither & 7 (1.3\%) & 16 (2.8\%) \\
Prefer not to answer & 3 (0.5\%) & 7 (1.2\%) \\
\bottomrule
\end{tabular}
\end{table}
\end{landscape}

\end{document}